\documentclass{sig-alternate}
\usepackage{mathptmx} 

\usepackage{fancyhdr}
\usepackage[normalem]{ulem}
\usepackage[hyphens]{url}
\usepackage[sort,nocompress]{cite}
\usepackage[final]{microtype}
\usepackage[keeplastbox]{flushend}
\usepackage[bookmarks=true,breaklinks=true,letterpaper=true,colorlinks,linkcolor=black,citecolor=blue,urlcolor=black]{hyperref}
\usepackage{threeparttable}
\usepackage{xcolor}
\usepackage{upgreek}
\usepackage{diagbox}

\newcommand{\instbit}[1]{\mbox{\scriptsize #1}}

\usepackage{hyperref}

\usepackage{pifont}
\usepackage{array}
\usepackage{makecell}
\usepackage{multicol}
\usepackage{epstopdf}
\usepackage{multirow}
\usepackage{framed}
\usepackage{authblk}

\fancypagestyle{firstpage}{
  \fancyhf{}

}

\usepackage[T1]{fontenc}
\usepackage[utf8]{inputenc}
\usepackage{authblk}

\title{\spaceskip=0.2em\relax{Domino: A Tailored Network-on-Chip Architecture to Enable Highly Localized Inter- and Intra-Memory DNN Computing}}
\author[*]{Kaining Zhou}
\author[*]{Yangshuo He}
\author[*]{Rui Xiao}
\author[*]{Kejie Huang}
\affil[*]{Zhejiang University}

\begin{document}
\maketitle
\thispagestyle{firstpage}
\pagestyle{plain}

\begin{abstract}
The ever-increasing computation complexity of fast-growing Deep Neural Networks (DNNs) has requested new computing paradigms to overcome the memory wall in conventional Von Neumann computing architectures. The emerging Computing-In-Memory (CIM) architecture has been a promising candidate to accelerate neural network computing. However, the data movement between CIM arrays may still dominate the total power consumption in conventional designs. This paper proposes a flexible CIM processor architecture named Domino to enable stream computing and local data access to significantly reduce the data movement energy. Meanwhile, Domino employs tailored distributed instruction scheduling within Network-on-Chip (NoC) to implement inter-memory-computing and attain mapping flexibility. The evaluation with prevailing CNN models shows that Domino achieves 1.15-to-9.49$\times$ power efficiency over several state-of-the-art CIM accelerators and improves the throughput by 1.57-to-12.96$\times$.

\end{abstract}

\section{Introduction}
 
The rapid development of Deep Neural Network (DNN) algorithms has led to high energy consumption due to millions of parameters and billions of operations in one inference\cite{vggnet,senet,Real_Aggarwal_Huang_Le_2019}. Meanwhile, the increasing demand for Artificial Intelligence (AI) computing entails flexible and power-efficient computing platforms to reduce inference energy and accelerate DNN processing. However, shrinkage in Complementary Metal-Oxide Semiconductor (CMOS) technology nodes abiding by Moore's Law has been near the end. Meanwhile, the conventional Von Neuman architectures encounter a ``memory wall'' where the delay and power dissipation of accessing data has been much higher than that of an Arithmetic Logic Unit (ALU), which is caused by the separation of storage and computing components in physical space. Therefore, new computation paradigms are pressed for the computing power demand for AI devices in the post-Moore's Law era.

One of the most promising solutions is to adopt Computing-in-Memory (CIM) scheme to greatly increase the parallel computation speed with much lower computation power. Recently, both volatile memory and non-volatile memory have been proposed as computing memories for CIM. SRAM is a volatile memory that enables partially digital \cite{ISSCC_2021_Jia,ISSCC_2021_Yue,ISSCC_2021_Chen,ISSCC_2021_Guo} or fully digital \cite{FloatPIM,isscc-digital-pim} CIM schemes and allows weight updating during inference. Resistive Random Access Memory (ReRAM), which has shown great advantages of high density, high resistance ratio, and low reading power, is one of the most promising candidates for CIM schemes \cite{reram1,reram2,reram3,reram4,reram5}. 

However, most research focuses only on the design of the CIM array, which lacks a flexible top-level architecture for configuring the storage and computing units of DNNs. Due to the high write power of ReRAM, the weight updating should be minimized \cite{AtomLayer,ISAAC}. Therefore new flexible interconnect architectures and mapping strategies should be employed to meet the various requirements of DNNs while achieving high components utilization and energy efficiency. There are two main challenges to designing a low-power flexible CIM processor. The first challenge is that complicated 4-D tensors have to be mapped to 2-D CIM arrays. The conventional 4-D tensor flattening method has to duplicate data or weight, which increases the on-chip memory requirement. The second challenge comes from the flexibility requirement to support various neural networks. Therefore, partial-sums and feature maps are usually stored in the external memory, and an external processor is generally required to maintain complicated data sequences and computing flow.

Network-on-Chip (NoC) with high parallelism and scalability has attracted a lot of attention from industry and academia\cite{Slim-NoC}. In particular, NoC can optimize the process of computing DNN algorithms by organizing multiple cores uniformly under specified hardware architectures \cite{TrueNorth,Loihi,eyerissv1,Eyerissv2,NoC-based-DNN}. This paper proposes a tailored NoC architecture called Domino to enable highly localized inter- and intra-memory computing for DNN inference. Intra-memory computing is the same as the conventional CIM array to execute Multiplication-and-Accumulation (MAC) operations in memory. Inter-memory computing is that the rest of computing (partial sum addition, activation, and pooling) is performed in the network when data are moving between CIM arrays. Consequently, ``computing-on-the-move'' dataflow is proposed to maximize data locality and significantly reduce the energy of the data movement. The dataflow is controlled by distributed local instructions instead of an external/global controller or processor. A synchronization and weight duplication scheme is put forward to maximize parallel computing and throughput. The high-energy efficient CIM array is also proposed. It is worth noting that various CIM schemes can be adopted in our proposed Domino architecture. The contributions of this paper are as follows:

$\bullet$ We propose an NoC based CIM processor architecture with distributed local memory and dedicated routers for input feature maps and output feature maps, which enables flexible dataflow and reduces the data movement significantly. The evaluation results show that Domino improves power efficiency and throughput by more than 15\% and 57\%, respectively.

$\bullet$ We define a set of instructions for Domino. The distributed, static, and localized instruction schedules are pre-loaded to Domino to control specific actions in the runtime. Consequently, our scheme avoids the overhead of the instruction and address movement in the NoC.

$\bullet$ We design a ``computing-on-the-move'' dataflow that performs Convolution Neural Network (CNN) related operations with our proposed NoC and local instruction tables. MAC operations are completed in CIMs, and other associated computations like partial sum addition, pooling, and activation are executed in the network when data are moving between CIM arrays.

The rest of the paper is organized as follows: \autoref{background} introduces the background of CNNs, dataflow, and NoCs; \autoref{Opportunity} details the opportunity and innovation with respect to Domino; \autoref{architecture} describes the architecture and building blocks of Domino; \autoref{dataflow} illustrates the computation and dataflow model; \autoref{instruction} defines the instruction and execution; \autoref{evaluation} presents the evaluation setup, experimental results, and experimental comparison; \autoref{related} introduces related works; finally, \autoref{conclusion} concludes this work.

\section{Background}
\label{background}
High computation costs of DNNS have posed challenges to AI devices for power-efficient computing in a real-time environment. Conventional processors such as CPUs and GPUs are power-hungry devices and inefficient for AI computations. Therefore, accelerators that improve computing efficiency are under intensive development to meet the power requirement in the post Moore's Law era.

\subsection{CNN Basics}
An essential computation operation of a CNN is convolution. Within a convolution layer (CONV layer), 3-D input activations are stacked as an Input Feature Map (IFM) to be convolved with a 4-D filter. The value and shape of an Output Feature Map (OFM) are determined by convolution results and other configurations such as convolution stride and padding type.

\begin{scriptsize}
    \begin{table}[h!]
    \centering
    \begin{tabular}{|l|l|}
        \hline
        \textbf{Parameter} & \textbf{Description}\\
        \hline
        $H/W$ & IFM height / width \\
        \hline
        $C$ & Number of IFM / filter channels\\
        \hline
        $P$ & Padding size \\
        \hline
        $K$ & Convolution filter height / width\\
        \hline
        $K_p$ & Pooling filter height / width\\
        \hline
        $M$ & Number of filters / OFM channels \\
        \hline
        $E/F$ & OFM height / width \\
        \hline
        $S$ & Convolution stride \\
        \hline
        $S_p$ & Pooling stride \\
        \hline

    \end{tabular}
    \caption{Shape Parameters of a CNN.}
    \label{tab:cnn_parameters}
    \end{table}
\end{scriptsize}

Given parameters in \autoref{tab:cnn_parameters}, and let $O,\ I,\ W$, and $B$ denote the OFM tensor, IFM tensor, weight tensors (filters), and bias (linear) tensor, respectively, the total OFM can be calculated as
\begin{equation}
    \label{equ:conv}
    \begin{split}
        \mathbf{O}[m][x][y] =  \sum\limits_{c = 0}^{C-1}\sum\limits_{i = 0}^{K-1}&\sum\limits_{j = 0}^{K-1} \mathbf{I}[c][Sx + i][Sy + j] \\
         \times \mathbf{W}[m][c]&[i][j] +\mathbf{B}[m],\\ 
       0 \leq m <  M,\  0\leq & x < E,\ 0\leq y < F, \\
       E=\lfloor \frac{H + 2P -K+S}{S} \rfloor &,\  F=\lfloor\frac{W + 2P-K+S}{S} \rfloor.
    \end{split}
\end{equation}

Usually, a nonlinear activation is applied on the OFM followed by pooling to reduce the spatial resolution and avoid data variation and distortion. After a series of CONV and pooling operations, Fully Connected (FC) layers are applied and performed to classify target objects.

\subsection{Dataflow}
The dataflow rules how data are generated, calculated and transmitted on an interconnected multi-core chip under a given topology. General dataflow for a CNN accelerator can be categorized into four types: Weight Stationary (WS), Input Stationary (IS), Output Stationary (OS), and Row Stationary (RS), based on the taxonomy and terminology proposed in \cite{Efficient}.

Kwon et al. propose MEARI, i.e., a type of flexible dataflow mapping for DNN with reconfigurable interconnects\cite{MAERI,meari2}. Chen et al. present a row-stationary dataflow adapting to their spatial architecture designed for CNNs processing\cite{eyerissv1}. Based on that, \cite{Eyerissv2} is put forward further as a more hierarchical and flexible architecture. FlexFlow is another dataflow model dealing with parallel types mismatch between the computation and CNN workloads\cite{FlexFlow}. These works attempt to make the best advantages of computation parallelism, data reuse, and flexibility\cite{In-Memory-Data-Parallel,Optimizing-Weight-Mapping,dasm}. 

\subsection{Network-on-Chip}
Current CIM-based DNN accelerators use a bus-based H-tree interconnect\cite{PipeLayer,max2}, where most latency of each different type of CNN is spent on communication \cite{Latency-Optimized}. A bus has limited address space and is forced synchronized on a complex chip. In contrast, an NoC can span synchronous and asynchronous clock domains or use asynchronous logic that is not clock-bound, to improve a processor's scalability and power efficiency. 

Unlike simple bus connections, routers and Processing Elements (PEs) are linked according to the network topology, which makes a variety of dataflows on an NoC. Following NNs' characteristics of massive calculation and heavy traffic requirements, the NoC is a prospective solution to provide an adaptive architecture basis. Kwon et al. propose an NoC generator that generates customized networks for dataflow within a chip\cite{rethink-nocs}. Firuzan et al. present a reconfigurable NoC architecture of 3-D memory-in-logic DNN accelerator \cite{Reconfigurable-noc}. Chen et al. put forward Eyeriss v2, which achieves high performance with a hierarchical mesh structure \cite{Eyerissv2}.

\definecolor{shadecolor}{rgb}{1,1,0}

\section{Opportunities \& Innovations} 
Aimed at challenges mentioned in Section 1, we identify and point out opportunities lying in prior propositions. 
\label{Opportunity}
\subsection{Opportunities}
\textbf{Opportunity \#1.} The CIM schemes significantly reduce the weight movement. However, they may still have to access the global buffer or external memory to complete the operations such as tensor transformation. partial addition, activation, and pooling. Methods like image-to-column and systolic matrix need to duplicate input data, leading to additional storage requirement and low data reuse. Therefore, the power consumption may be still dominated by the data movement.

\textbf{Opportunity \#2.} Some overhead still exists in instruction and address transmission on an NoC. They need to be dispatched from a global controller, and transmitted in the router. These overheads will cause extra bandwidth requirement and bring in long latency and synchronization issues.

\subsection{Domino Innovations}
In face of the aforementioned opportunities and challenges, we employ the ``computing-on-the-move'' dataflow, localized static schedule table, together with tailored routers to avoid the cost of off-chip accessing, instruction transmitting over NoC. Following are our innovations in Domino's architecture.

\textbf{Innovation \#1.} Domino adopts CIM array to minimize the energy cost of weight refreshing and transmission. The data are transmitted through dedicated dual routers NoC structure, where one router is for input feature map processing and another one is equipped with a computation unit and schedule table for localized control and partial-sum addition. The localized data processing greatly reduces the energy for data movement.

\textbf{Innovation \#2.} Domino adopts ``computing-on-the-move'' dataflow to reduce the excessive and complex input and partial-sum movement and tensor transformation. In prior works of CIM-based architectures, the data have to be duplicated or stored in extra buffers for future reuses, the addition of partial-sum is either be performed in an external accumulator or a space-consuming adder tree outside PE arrays \cite{MAERI}. The ``computing-on-the-move'' dataflow used in Domino aims at performing the extra computation in data movements and maximize data locality. Domino reuses input by transferring over the array of tiles and adds partial-sums along unified routers on NoC. The overall computation can be completed on-chip and no off-chip data accesses are required. 

\textbf{Innovation \#3.} Domino adopts localized instructions in routers to achieve flexible and distributed control. We identify that existing designs of ``computing-on-the-move'' dataflow, like \cite{active-routing}, are supported by the external controller or the head information including source and destination in transmission packets. They introduce additional transmission delay, bandwidth requirement, and power consumption. A local instruction table enables distributed and self-controlled computation to reduce the energy and time consumed by external instruction or control signals. The instructions fit and support Domino's ``compute-on-the-move'' well.

\section{Domino Architecture}
\label{architecture}

This section details the designed architecture for DNN processing. To boost DNN computation while maintaining flexibility, we propose an architecture called Domino. The purpose of Domino's architecture is to enable ``computing-on-the-move'' within a chip, so Domino assigns hardware resources uniformly and identically on each building block of different hierarchy.

From a top view, Domino consists of an input buffer and $A_r\times A_c$ tiles interconnected in a 2-D mesh NoC. The number of tiles is adjusted for different applications. The weights and configuration parameters are loaded initially. The input buffer is used to store the required input data temporarily. A Domino block is an array of tiles virtually split in mesh NoC to serve a DNN layer. A tile contains a PE performing in-memory DNN computation, and the two routers transmit the results in a tile. By this means, Domino achieves a high level of distributed computation and uniformity, making it a hierarchical, flexible, and easily reconfigurable DNN processor architecture.

\subsection{Domino Block}
 
\begin{figure}[htbp]
    \centering  
    \includegraphics[width=0.4\textwidth]{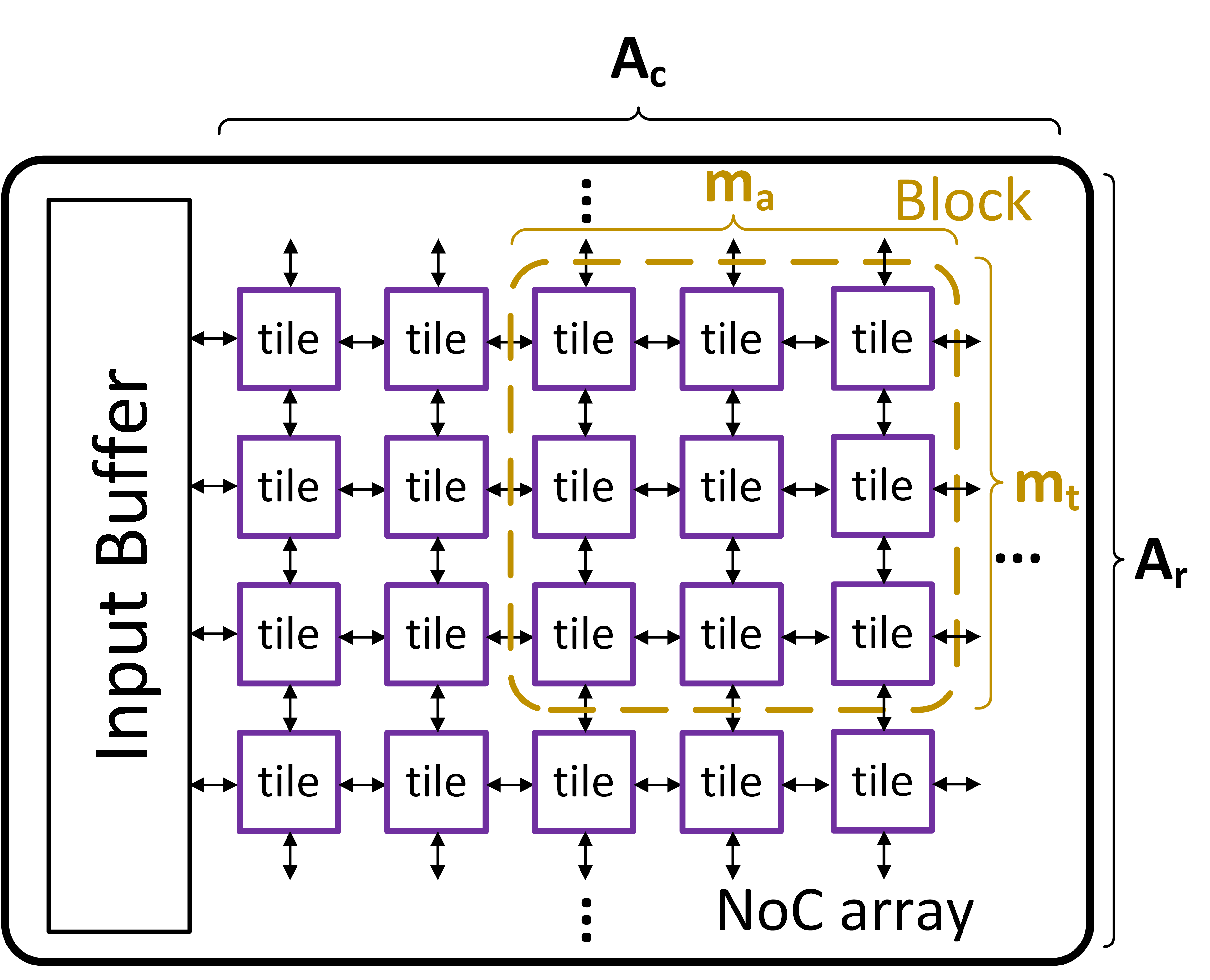}
    \caption{The top-level block diagram of the Domino architecture. A Domino block is a $m_t\times m_a$ array of tiles on NoC for computation of a DNN layer.}  
    \label{fig:block}
\end{figure}

Domino is virtually split into blocks corresponding to the layers in a neural network. Let $m_t$ and $m_a$ denote the number of rows and columns of a tile array in a block. In the initialization and configuration stages, a $m_t\times m_a$ array of tiles are assigned to form a block used to deal with a CONV layer together with a pooling layer (if needed) or an FC layer. A block provides interconnection via a bi-direction link between tiles in four directions, as shown in \autoref{fig:block}. A Domino block provides diverse organizations of tiles for various dataflows based on layer configurations, including DNN weight duplication and block reuse schemes. For instance, in pursuit of layer synchronization, block adopts weight duplication that the block processes $m_a$ rows of pixels simultaneously. We will discuss the details of block variations in the dataflow section.

\subsection{Domino Tile}

\begin{figure}[]
    \centering  
    \includegraphics[width=0.45\textwidth]{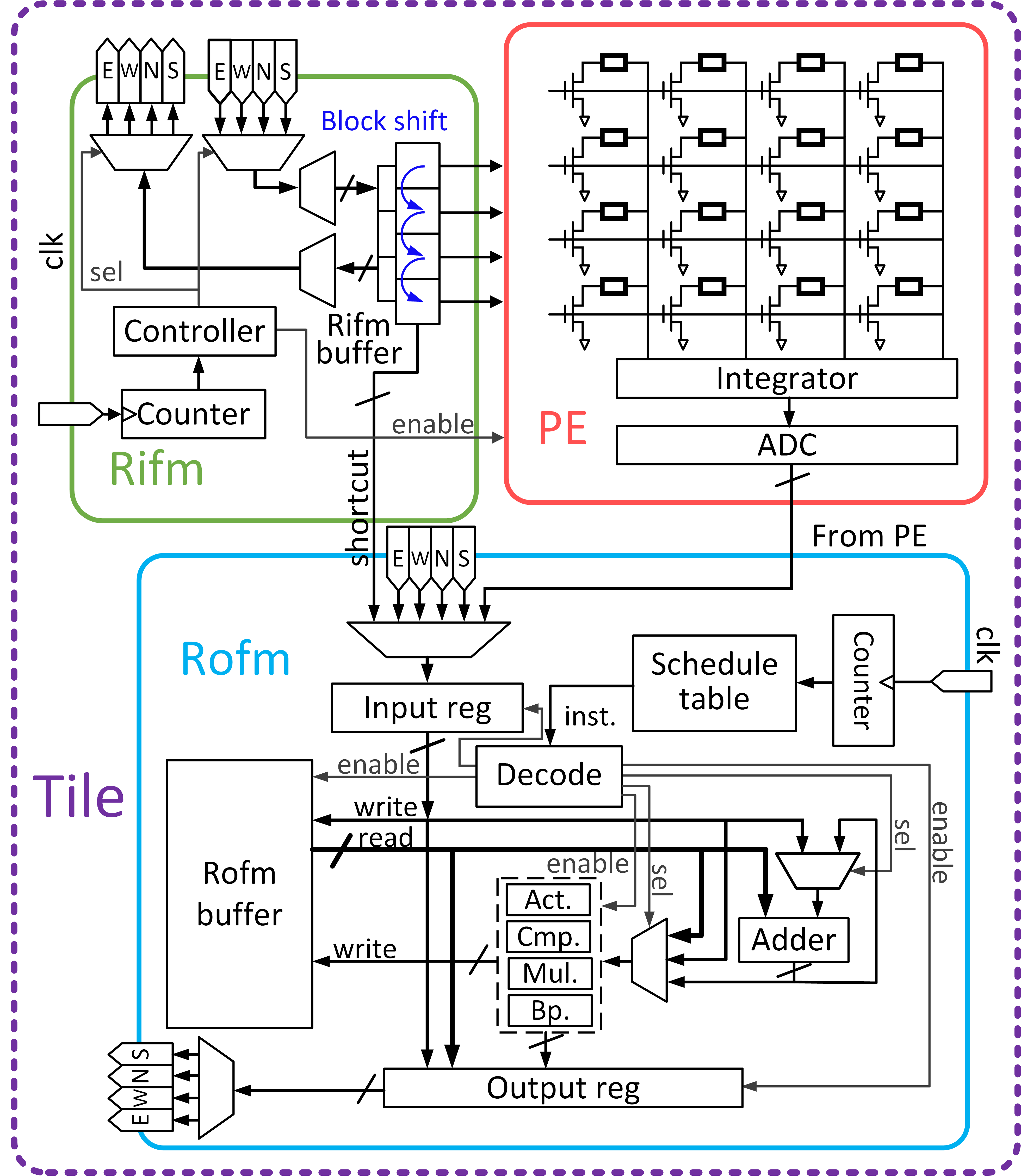}
    \caption{A Domino tile contains two routers Rifm and Rofm and a computation center PE.}  
    \label{fig:tile}
\end{figure}

A tile is the primary component of Domino that is used for DNN computation. It involves a CIM array called PE, a router transferring IFMs called Rifm, and a router transferring OFMs or their partial-sums in convolution computation called Rofm. The basic structure of a tile is illustrated in \autoref{fig:tile}. Rifm receives input data from one out of four directions in each tile and controls the input dataflow to remote Rifm, local PE, and local Rofm. The in-memory computing starts from the Rifm buffer and ends at Analog-to-Digital (ADC) converters in PE. The outputs of PE are sent to Rofm for temporary storage or partial-sum addition. Rofm is controlled by a series of periodic instructions to receive either computation results or input data via a shortcut from Rifm and maintain the dataflow to adding up partial-sums.

\subsection{Domino Rifm}

\label{sec:Rifm}
As shown in \autoref{fig:tile}, each Rifm possesses I/O ports in four directions to communicate with Rifms in the adjacent tiles. It is also equipped with a buffer called Rifm buffer to store received input data in the current cycle. Moreover, Rifm has in-tile connections with PE and Rofm. The bits in the Rifm buffer controlled by an 8-to-1 MUX are sequentially sent to PE for MAC computations. The buffer size is 8$\times N_c$ bits where $N_c$ is the number of rows in PE (PE consists a $N_c\times N_m$ ReRAM array which is discussed in \autoref{sec:PE}). It supports an in-buffer shifting operation with a step size of $64$ or the multiple of $64$, which supports the case when the input channels of a layer are less than $N_c$. A shortcut connection from Rifm to Rofm is established to support the situation that MAC computation is skipped (i.e., the shortcut in a ResUnit). A counter and a controller in Rifm decide input dataflow based on the initial configuration. Once Rifm receives input packets, the counter starts to increment the counter's value. The controller chooses to activate MAC computation or sends ``enable" signals to I/O ports to receive or transmit data based on the initial configuration and the counter's value.

\subsection{Domino Rofm}
Rofm is the key component for ``computing-on-the-move'' dataflow controlled by instructions to manage I/O ports and buffers, add up partial/group-sum results, and perform activation or pooling to get convolution results. \autoref{fig:tile} shows the micro-architecture in Rofm, which consists of a set of four-direction I/O ports, input/output registers, an instruction schedule table, a counter to generate instruction indices, a Rofm buffer to store partial computation results, a reusable adder, a computation unit with adequate functions, and a decoder. The instructions are generated by the compiler based on DNN configurations. The instructions in the schedule table are executed periodically based on initial configurations. The internal counter starts functioning and keeps increasing its value as soon as any input packet is received. The decoder reads in the value of counter every cycle used as the index of the schedule table to fetch an instruction. Instructions are then decoded and split into control words to control ports, buffers, and other computation circuits within Rofm. The partial-sums are added to group-sums when transferring between tiles. The group-sums are queued in the buffer for other group-sums to be ready and then form a complete computation result.

A computation unit is equipped in each Rofm to cope with the non-linear operations in DNN computation, including activation and pooling, as shown in \autoref{fig:tile}. Activation is only used in the last tile of a block. The comparator is used for the max-pooling, which outputs the larger value of the convolution results from adjacent Rofms. The multiplier and adder perform the average-pooling.

\begin{figure}[htbp]
    \centering  
    \includegraphics[width=0.48\textwidth]{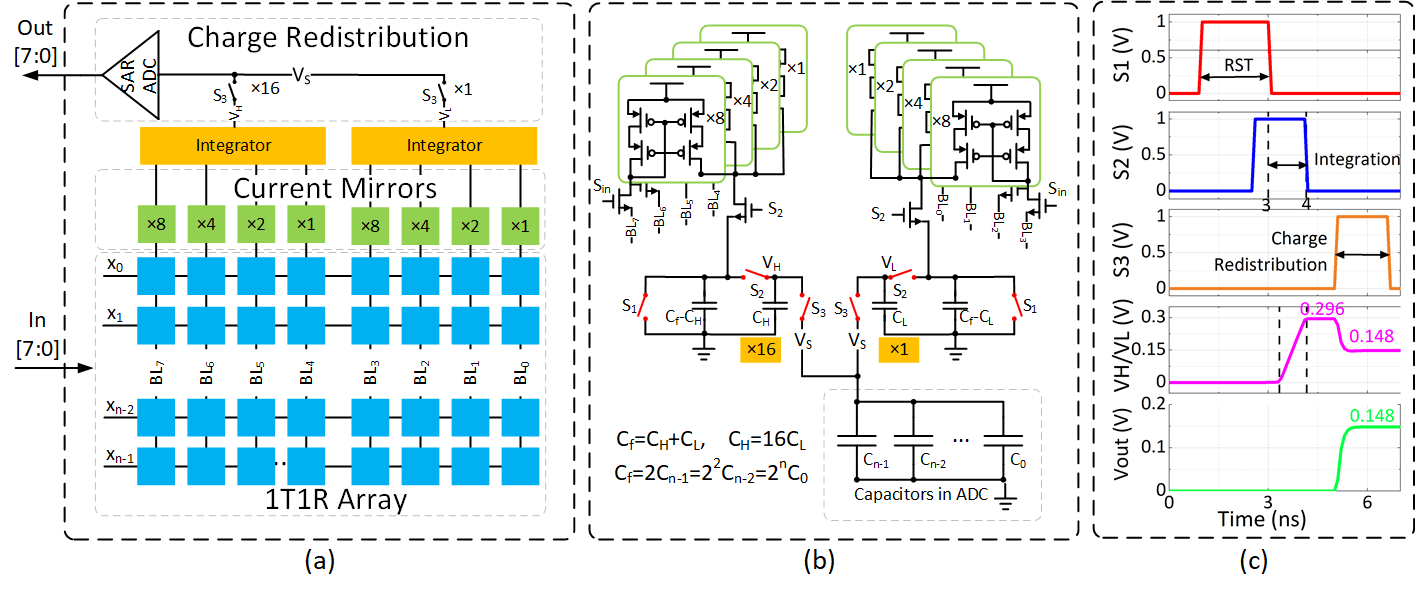}
    \caption{Domino PE: (a) The block diagram of the Domino PE, (b) the neuron circuit in the Domino PE, and (c) the simulation waveform of the PE.}  
    \label{fig:CIM_Array}
\end{figure}

\subsection{Domino PE}
\label{sec:PE}
PE is the computation center that links a Rifm and a Rofm in the same tile. \autoref{fig:CIM_Array} (a) shows the architecture of the proposed Domino PE, which is composed of a 1T1R crossbar array, current mirrors, two integrators, and a Successive Approximation Register (SAR) ADC \cite{zhang2020regulator}. As shown in \autoref{fig:CIM_Array} (b), eight single-level 1T1R cells are used to represent one 8-bit weight $w_{[7:0]}$. The voltage on ReRAM is clamped by the gate voltage and threshold voltage, which will generate two levels' of current based on the status of ReRAM and access transistors. Weight bits are divided into two sets ($BL_{7-4}$ and $BL_{3-0}$). In each set, four current mirrors are used to provide the significance for each bit line ($\frac{k}{8}$, $\frac{k}{4}$, $\frac{k}{2}$, and $k$, where $k$ is the gain to control the integration speed). The output of current mirrors is accumulated in the integrator. The higher four bits and lower four bits are jointed by the charge redistribution between two capacitors in the two integrators with a ratio of 16:1. The significance of the input data is also realized by averaging the charge between integrators and ADC \cite{zhang2020regulator}. The simulation waveform of the integration and charge sharing process is shown in \autoref{fig:CIM_Array} (c). In the initial configuration, neural network weights are mapped to ReRAM arrays in PEs, which will be discussed in detail in \autoref{dataflow}. 

\section{Dataflow Model}
\label{dataflow}

We propose ``computing-on-the-move'' dataflow to reduce both data movement and data duplication. Below we will introduce the dataflow in FC layers, CONV layers, and pooling layers, then reveal how it supports different types of DNN and achieve high performances. The other ``computing-on-the-way'' schemes such as MAERI\cite{MAERI} and Active-routing\cite{active-routing} map computation elements to memory network for near-memory computing and aggregate intermediate results when transmitting in a tree adder controlled by a host CPU. In contrast, Domino conducts MAC operations in memory. The partial-sums are added up to get group-sums, which are then stored in the buffer to wait for other group-sums ready for addition. The group-sums are added up when moving along the routers controlled by local schedule tables. Domino only transmits data while MAERI/Active-routing contains extra complex information such as operator and flow states.

\subsection{Dataflow in FC layers}
\begin{figure}[ht]
    \centering  
    \includegraphics[width=0.4\textwidth]{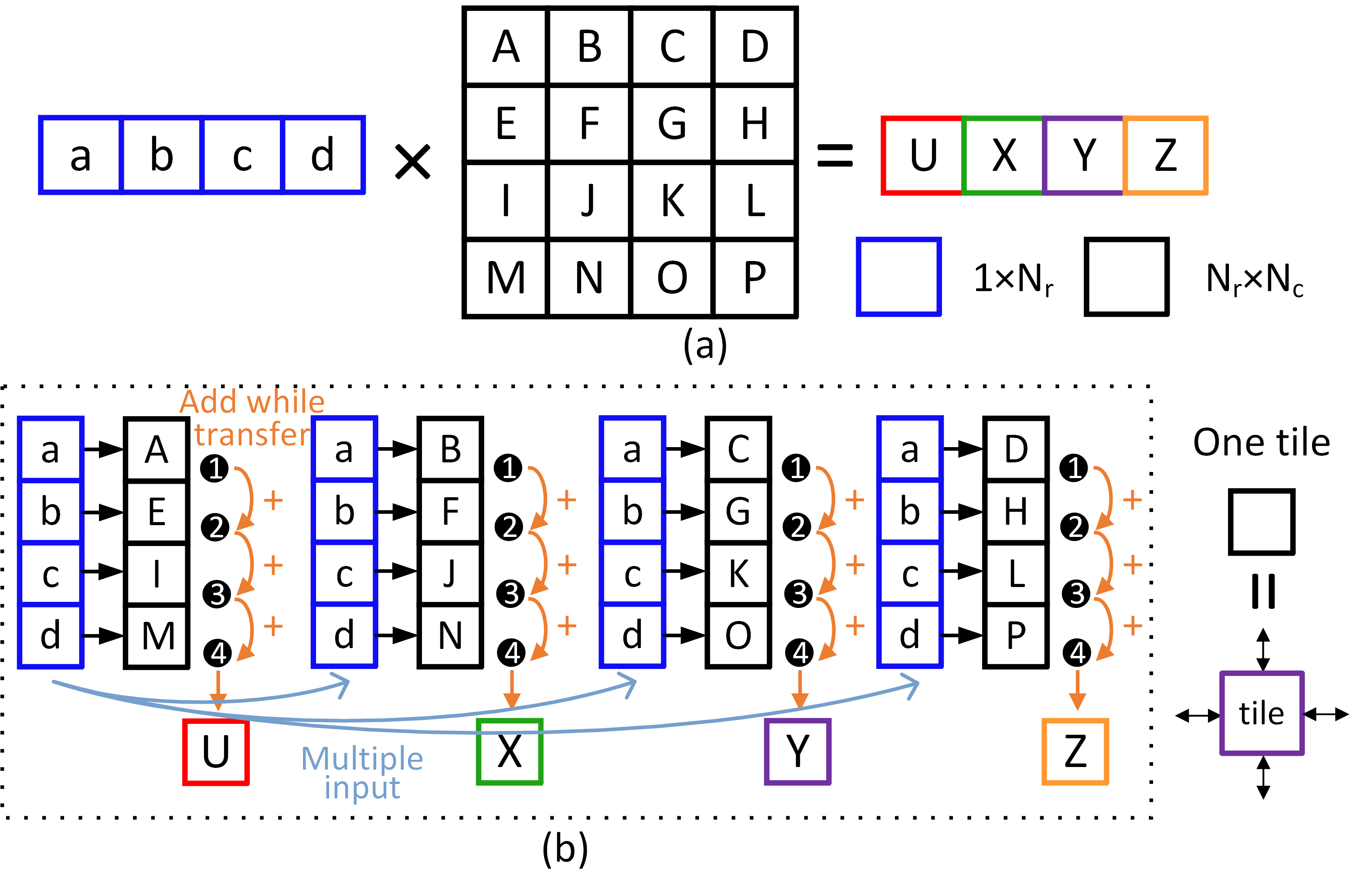}
    \caption{The proposed mapping and dataflow in FC layers: (a) the partitioned input vector and weight matrix; (b) the dataflow to transmit and add multiplication results to a complete MVM result.}  
    \label{fig:fc_dataflow}
\end{figure}

In an FC layer, IFMs are flattened to a vector with dimension $1\times C_{in}$ to perform a Matrix-Vector Multiplication (MVM) computing with the weights. MVM can be formulated as $\mathbf{y} = \mathbf{xW}$, where $\mathbf{x}\in  \mathbb{R}^{1\times C_{in}}$, $\mathbf{y}\in  \mathbb{R}^{1\times C_{out}}$ are input and output vector, respectively, and $\mathbf{W}\in \mathbb{R}^{C_{in}\times C_{out}}$ is a weight matrix.

In most cases, the input and output vector dimensions in FC layers are larger than the size of a single crossbar array in PE. Therefore, the partitioned matrix multiplication as described in \autoref{equ:FC_layer_divided} is used to deal with the above-mentioned issue: the $C_{in}\times C_{out}$ weight matrix is divided into $m_t\times m_a$ smaller matrices $\mathbf{W}_{ij}$ with the dimension of $N_c\times N_m$. The input vector is divided into $m_t$ small slices $\mathbf{x}_i$ to multiply the partitioned weight matrices. As shown in \autoref{fig:fc_dataflow} (a), the sum of the multiplication results within a column of partitioned matrices is a slice of output vector $\mathbf{y}_j$. The slices of vectors are concatenated to produce the complete output. 

\begin{equation}
    \label{equ:FC_layer_divided}
    \begin{split}
        \mathbf{x} & = [\mathbf{x}_1,\mathbf{x}_2\cdots,\mathbf{x}_{m_t}] \\
        \mathbf{W} & = \left[\begin{array}{cccc}
            \mathbf{W}_{11} & \mathbf{W}_{12} & \cdots & \mathbf{W}_{1m_a}  \\
            \vdots & \vdots & \ddots & \vdots \\
            \mathbf{W}_{m_t1} & \mathbf{W}_{m_t2} & \cdots & \mathbf{W}_{m_tm_a}  \\
        \end{array}\right]\\
        \mathbf{y} & = [\mathbf{y}_1,\mathbf{y}_2\cdots,\mathbf{y}_{m_a}],\quad \mathbf{y}_j = \sum\limits_{i=1}^{m_t}\mathbf{x}_i\mathbf{W_{ij}} \\
    \end{split}
\end{equation}

We propose a mapping scheme for efficient computation in FC layers to efficiently handle partitioned matrix multiplication in our tile-based Domino. As shown in \autoref{fig:fc_dataflow} (b), the partitioned blocks are mapped to $m_t\times m_a$ tiles ($m_t = \lceil\frac{C_{in}}{N_c}\rceil,\ m_a = \lceil\frac{C_{out}}{N_m}\rceil$), which is divided into four columns and four rows. The input vectors are transmitted to all $m_a$ columns of tiles. The multiplication results, \ding{182} to \ding{185}, are added while transmitting along the column. The final addition results in the last tiles of the four columns, $\mathbf{U}$ to $\mathbf{Z}$, are small slices of an output vector. Concatenating the small slices in all columns gives the complete partitioned matrix multiplication result.

\subsection{Dataflow in CONV Layers}

\begin{figure}[ht]
    \centering  
    \includegraphics[width=0.48\textwidth]{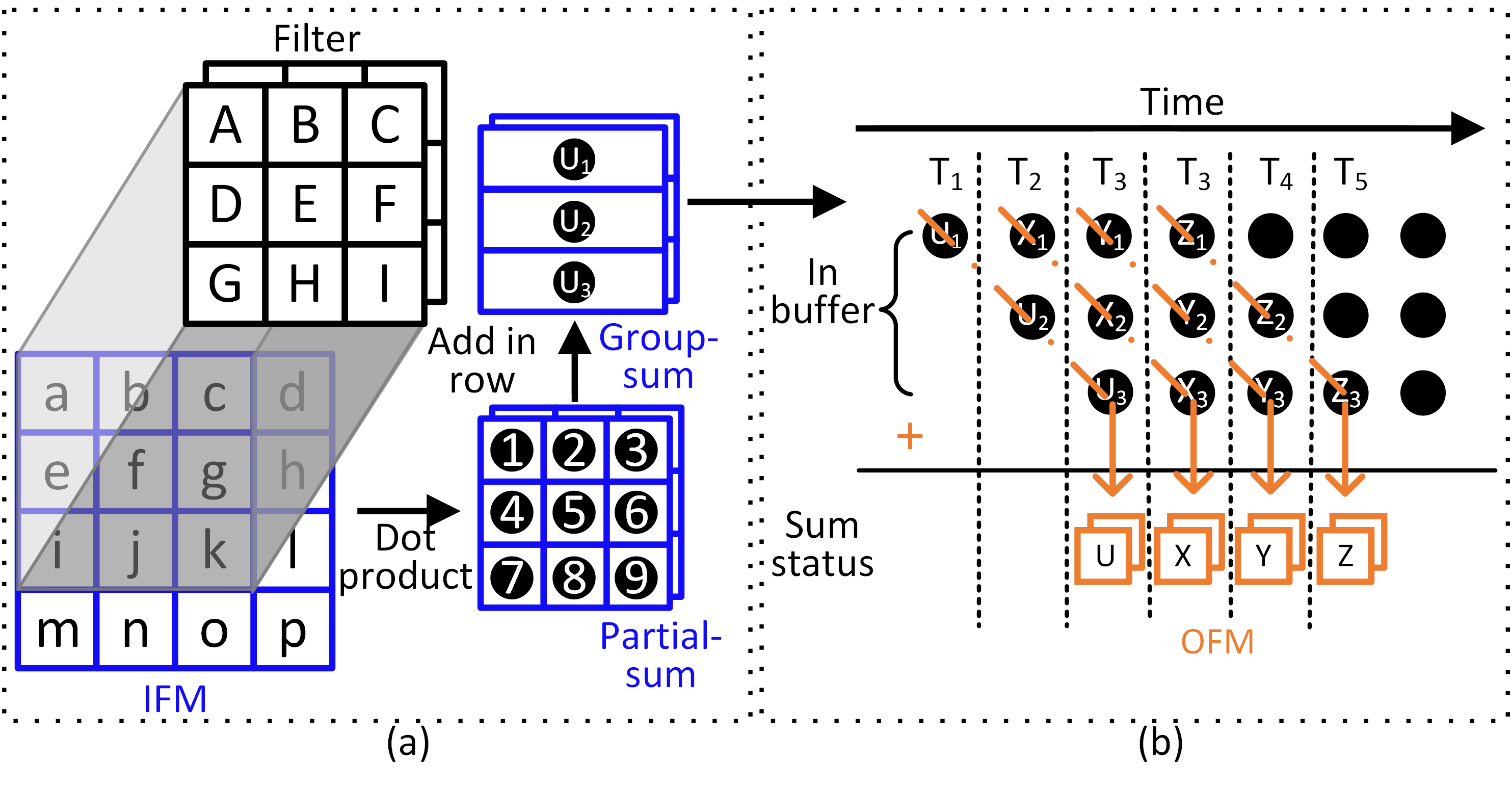}
    \caption{(a) The defined partial-sum and group-sum; (b) The timing and location of how Domino generates OFM with naive dataflow.}  
    \label{fig:partial_group_sum}
\end{figure}

Based on \autoref{equ:conv}, the computation of a pixel in OFM is the sum of point-wise MAC results in a sliding window. As shown in \autoref{fig:partial_group_sum}, we define the $N_m$ point-wise and row-wise MAC results as the partial-sum and group-sum, respectively. This figure illustrates our distributed way of adding partial-sums and timing control of ``computing on the move'' dataflow. Let $P_a^{(l)}$ denote the partial-sum of the $l^{\text{th}}$ pixels in the $a^{\text{th}}$ sliding window and $G_a^{(b)}$ denote the group-sum of the $b^{\text{th}}$ row in the $a^{\text{th}}$ sliding window.   
\begin{equation}
    \label{equ:group_sum}
    \begin{split}
      P_a^{(l)} = \sum\limits_{c=0}^{C-1}\mathbf{I}[Sx+i]&[Sy+j][c]\times \mathbf{W}[i][j][c] \\
      G_a^{(b)} = \sum\limits_{c=0}^{C-1}\sum\limits_{j=0}^{c-1}\mathbf{I}[Sx+b]&[Sy+j][c]\times \mathbf{W}[b][j][c]\\
        \mathbf{O}[x][y] = & \sum\limits_{b=0}^{K-1}G_a^{(b)}\\
    \end{split}
\end{equation}
where $l = iK+j$, and $a = xF + y$. $P_a^{(l)},\ G_a^{(b)},\ \mathbf{O}[x][y]$, and $\mathbf{W}[i][j][c] \in\mathbb{R}^{M}$. As shown in \autoref{fig:partial_group_sum} (a), $K$ partial-sums (\ding{182}, \ding{183}, and \ding{184}) can be added up to be a group-sum ($\mathbf{U_1}$). \autoref{fig:partial_group_sum} (b) illustrates how $K$ group-sums ($\mathbf{U_1}$, $\mathbf{U_2}$, and $\mathbf{U_3}$) are added up to be a complete convolution result $\mathbf{U}$ (the same for $\mathbf{X}$, $\mathbf{Y}$, $\mathbf{Z}$). $\mathbf{U_1}$, $\mathbf{U_2}$, and $\mathbf{U_3}$ are sequentially generated and summed up one by one, in different timing and tiles. The group-sums wait in the buffer for ready of other group-sum, or are evicted once they are no longer needed. 

\begin{figure}[ht]
    \centering  
    \includegraphics[width=0.48\textwidth]{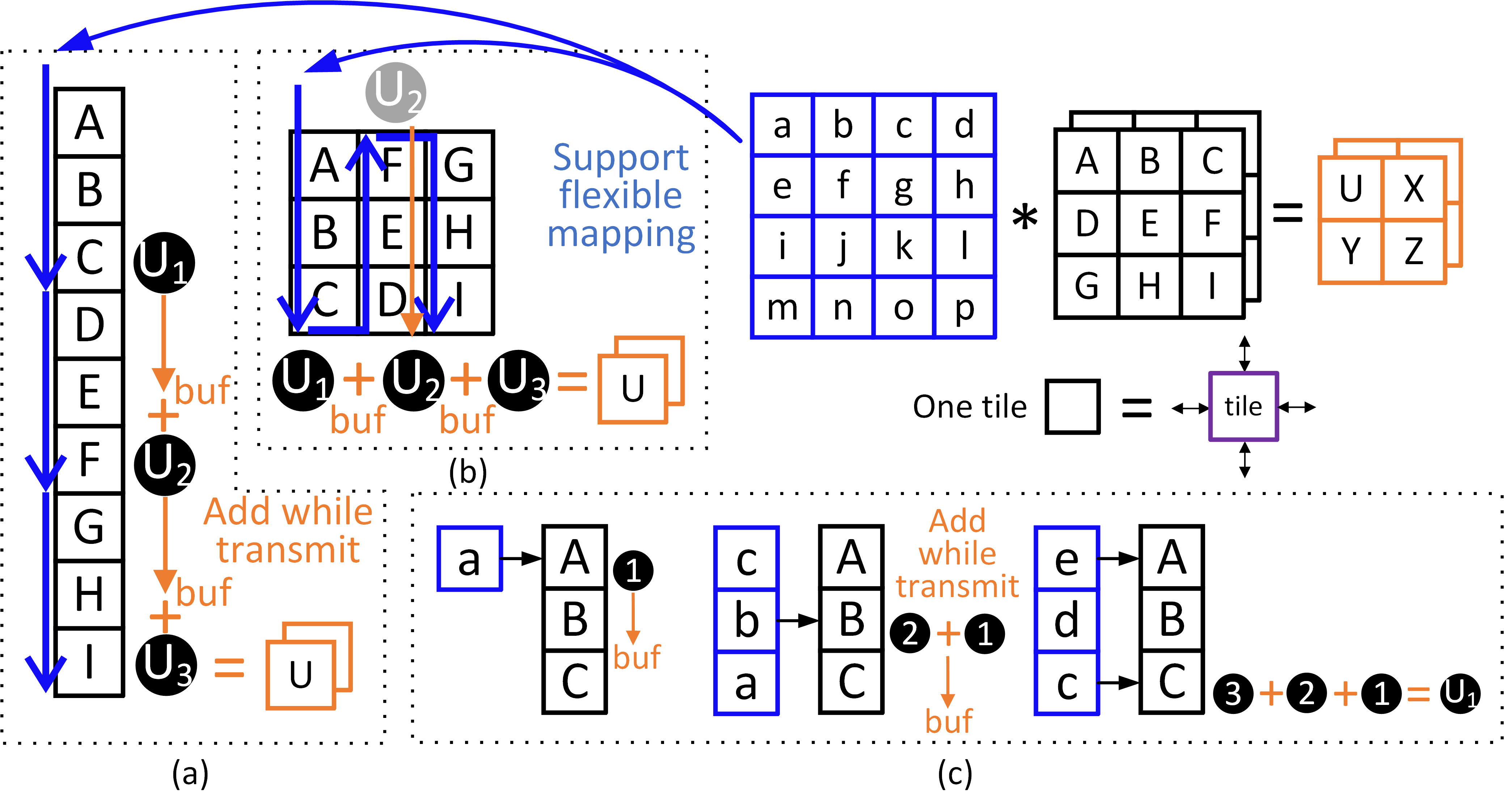}
    \caption{The proposed mapping and dataflow in CONV layer: (a) a $K^2\times 1$ tile array mapping weights in a CONV layer; (b) another type of $K\times K$ tiles array mapping weights in a CONV layer; (c) adding a group-sum while transmitting MAC results in a group of tiles.}  
    \label{fig:conv_dataflow_nodup}
\end{figure}

As shown in \autoref{fig:conv_dataflow_nodup} (a), the input channels and output channels of a CONV layer are mapped to the inputs and the outputs of a tile, respectively. If $N_c = C$ and $N_m = M$, $K^2$ points of the filters are mapped to $K^2$ tiles. In the case that the dimension of a weight matrix exceeds the size of the ReRAM crossbar array in PE ($N_c \leq C$ and $N_m \leq M$), the filters are split and mapped to $\lceil\frac{C}{N_c}\rceil\times\lceil\frac{M}{N_m}\rceil$ tiles. The tiles are placed closely to minimize the data transmission. Likewise, the input vector is split and contained in $\lceil\frac{C}{N_c}\rceil$ input packets. Adding and concatenating MAC results of these PEs will generate the same result as the non-split case. This case is similar to the FC dataflow as discussed earlier. When $N_c > C$, multiple points in a filter can be mapped to the same tile to improve ReRAM cells' utilization and reduce the energy for data movement and partial-sum addition. In such a scenario, the CONV computing is accomplished by the in-buffer shifting operation. In case $N_m \geq 2M$, the filters can be duplicated inside a tile to maximize parallel computing.

As shown in \autoref{fig:conv_dataflow_nodup} (a) and (b), each split block is allocated with an array of tiles that maps weights in a CONV layer. The array has multiple mapping typologies, such as $K^2\times  1$ ($m_t = K^2,\ m_a = 1$) and $K\times K$ ($m_t = m_a = K$), in pursuit of flexible dataflow and full utilization of tiles in the NoC array. As defined in \autoref{equ:group_sum}, a group-sum is the sum of $K$ partial-sums. We can cluster $K$ tiles mapped into the same row to a group. We categorize the CONV dataflow into three types: input dataflow, partial-sum dataflow, and group-sum dataflow. 

\textbf{The input dataflow}. The blue arrows in \autoref{fig:conv_dataflow_nodup} (a) illustrate the input dataflow in a $K^2\times 1$ array of tiles. The inputs are transmitted to an array in rows and flow through Rifms in $K^2$ tiles with identical I/O directions. Another type of zigzag dataflow is demonstrated in \autoref{fig:conv_dataflow_nodup} (b) in a $K\times K$ array of tiles. When the input data reach the last tile of a group, Rifm inverts the flow direction, and the input data are transmitted to the tile at the adjacent column.

\textbf{The partial-sum dataflow}. The partial-sum dataflow is shown in \autoref{fig:conv_dataflow_nodup} (c). $\mathbf{a}$ represents the first input data of IFM. At each cycle, input packets are transmitted to the Rifm of the succeeding tile. Meanwhile, MAC computation in PE is enabled if needed. Take input data $\mathbf{a}$ as an example, in the first cycle, $\mathbf{a}$ is multiplied with weight $\mathbf{A}$ to produce a partial-sum \ding{182} which will be transmitted by Rofm to the next tile. In the second cycle, $\mathbf{a}$ is transmitted to tile whose PE is mapped with weight $\mathbf{B}$. The controller in the Rifm will bypass MAC computation because $\mathbf{a}$ only multiplies with $\mathbf{A}$ in convolution. In the meanwhile, \ding{182} is stored in the Rofm buffer waiting for addition. In the third cycle, $\mathbf{b}$ is multiplied with $\mathbf{B}$ in PE, and partial-sum \ding{183} is received by Rofm. Then \ding{182} is popped out from the Rofm buffer and added with \ding{183}. Generally, partial-sums are added up to be a group-sum $\mathbf{U_1}$ when transmitting along a group of tiles.

\textbf{The group-sum dataflow}. Adding group-sums to convolution results is also a process executed during data transmission. As shown in \autoref{fig:conv_dataflow_nodup} (a) and (b), the orange arrows indicate the group-sum dataflow. Group-sums add up to get the convolution result $\mathbf{U}$ in the last tile of each group. When $\mathbf{U_1}$ is generated, it will wait in the third tile until $\mathbf{U_2}$ is generated in the sixth tile. Then, $\mathbf{U_1}$ will be transferred to the sixth tile to add $\mathbf{U_2}$. Similarly, $\mathbf{U_1+U_2}$ waits in the sixth tile for $\mathbf{U_3}$ to generate a complete result $\mathbf{U}$. After all linear matrix computation, the computation unit in Rofm controlled by instructions takes effect. An activation function is applied on the complete convolution result in Rofm in the last tile.

\subsection{Synchronization}
\label{sec:sync}

\begin{figure}[ht]
    \centering  
    \includegraphics[width=0.44\textwidth]{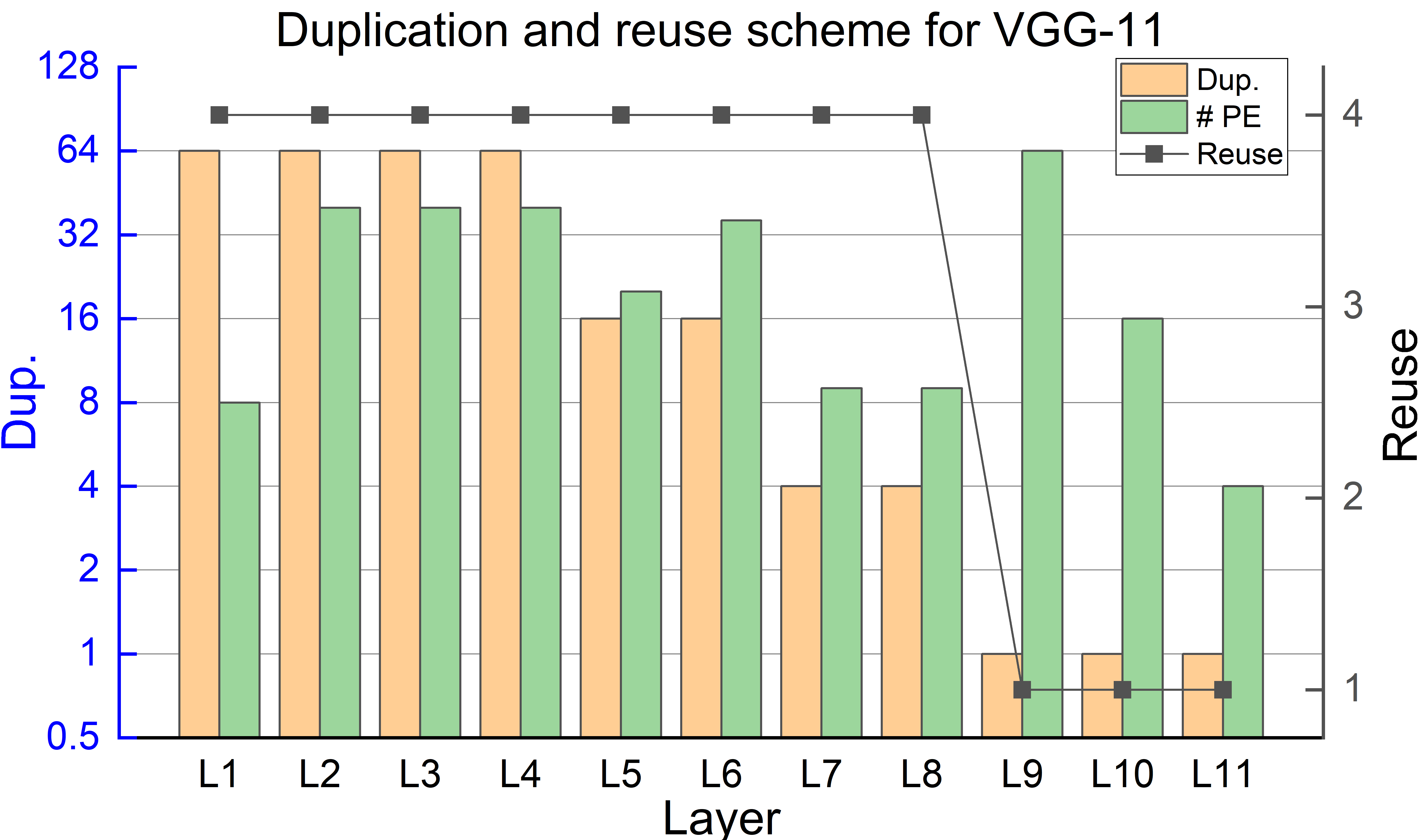}
    \caption{Duplication and reuse scheme in VGG-11 model used in \cite{ISSCC_2021_Jia}, there are three pooling layers before L5, L7, and L9. The left axis shows the number of tiles and duplication, and the right axis shows the number of reuses.}  
    \label{fig:Dup_Reuse}
\end{figure}

Because of down sampling in CONV layers, the CONV steps change with layers. For example, if the pooling filter size $K_p = 2$ and the pooling filter stride $S_p = 2$, every four OFM pixels will produce a pooling result. Therefore, the computing speed of the next layer has to be four times slower than the preceding layer, which will waste the hardware resource and severely affect the computing throughput. To maximize computation parallelism and throughput, the CONV filters need to be duplicated. In other words, the weights of the preceding filters should be duplicated by four times in the above example. 

However, a DNN model usually has a few down sampling layers. As a result, the number of duplicated tiles may exceed the total number of tiles in Domino. Therefore, a block reuse scheme is proposed for such a situation to alleviate heavy hardware requirements. \autoref{fig:Dup_Reuse} demonstrates our weight duplication and block reuse scheme for VGG-11 model used in \cite{ISSCC_2021_Jia} (CIFAR-10 dataset). There are three pooling layers, and the CONV steps before these three layers are 64, 16, and 4. If all layers are synchronized, a total of 892 tiles are required to map the network. If we reuse the tiles for all CONV layers by four times, the speed of the CONV layers will be four times faster than the FC layers. In such a situation, a total of 286 tiles are required to map the network. It can be concluded that there is a trade-off between chip size and throughput.

\subsection{Dataflow with Weight Duplication}

\begin{figure}[ht]
    \centering  
    \includegraphics[width=0.45\textwidth]{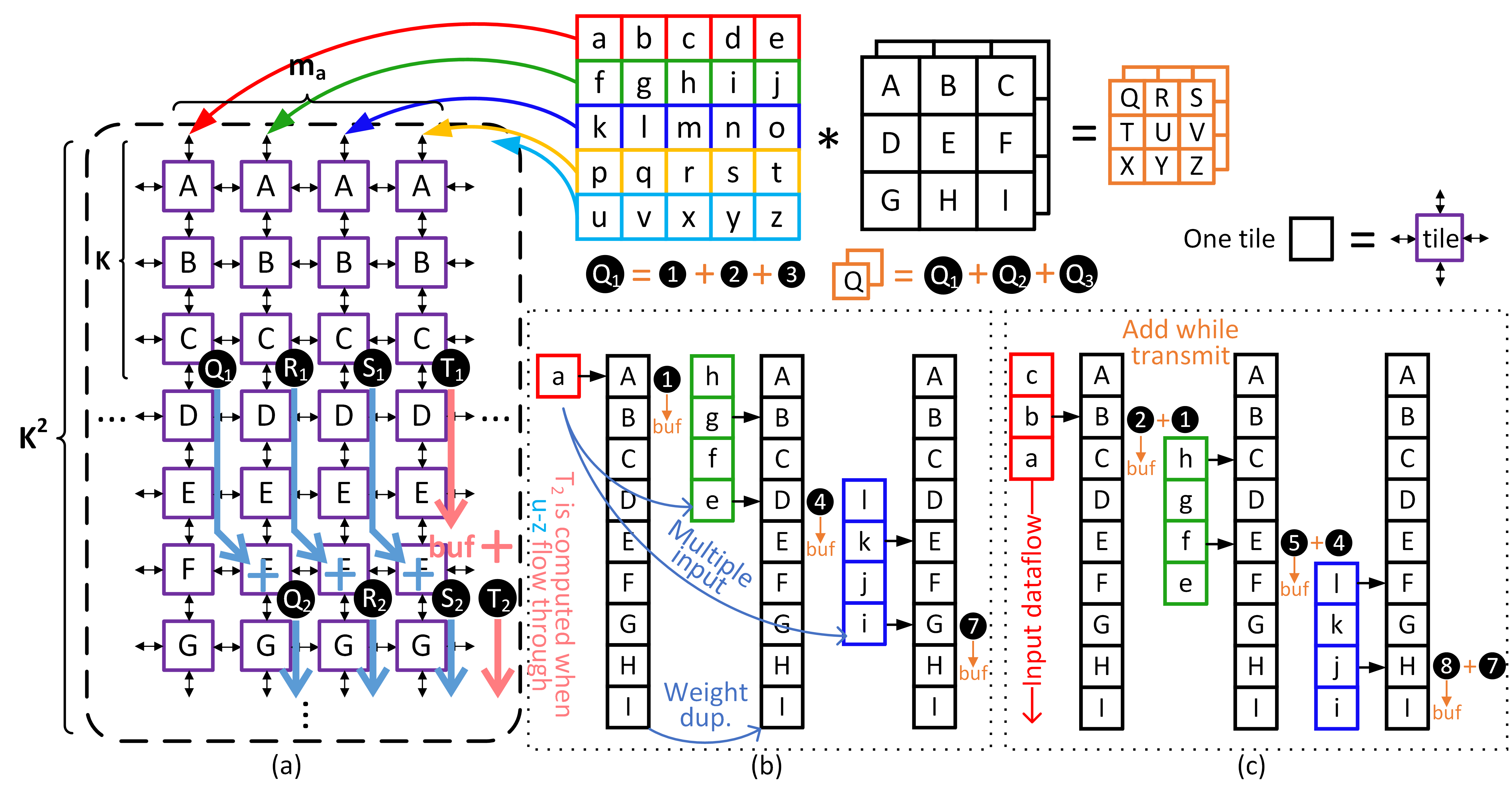}
    \caption{The proposed mapping and dataflow in CONV layers with weight duplication: (a) each column of tiles is assigned with duplicated weights and receives different rows of data in IFM; (b) (c) adding partial-sums to be group-sums while transmitting at two cycles.}  
    \label{fig:conv_dataflow_dup}
\end{figure}

Take weight duplication and reuse into consideration, the dataflow should be modified to match such scheme. The overall dataflow with weight duplication is illustrated in \autoref{fig:conv_dataflow_dup}. In this scenario, a block has $m_a$ ($m_a=4$ is the number of duplication in this example) duplicated $K^2\times 1$ ($m_t=K^2$) arrays of tiles.

\textbf{The input dataflow}. The input dataflow is demonstrated in \autoref{fig:conv_dataflow_dup} (a): four rows of data of an IFM are transmitted through four arrays of tiles in parallel. The massive data transmission is alleviated by leveraging the spatial locality. Every $m_a$ rows of data are alternatively transferred to the tiles in reverse order. As shown in \autoref{fig:conv_dataflow_dup} (a), the first four rows are transmitted to the block in increasing order. The fifth row of an IFM flows into the last column of the block.
Group-sum $\mathbf{T_1}$ is computed in the third tile when the fourth row flows through the tile. When the fifth row flows through, group-sum $\mathbf{T_2}$ is computed in the sixth tile. Since $\mathbf{T_1}$ and $\mathbf{T_2}$ are generated in the same column of a block, inter-column data transfer is reduced.

\textbf{The partial-sum dataflow}. \autoref{fig:conv_dataflow_dup} (b) and (c) depict the partial-sum dataflow at a certain cycle and the third cycle thereafter, respectively. In \autoref{fig:conv_dataflow_dup} (b), partial-sums \ding{182}, \ding{185}, and \ding{188} are computed in the first tiles of their groups and stored in the Rofm buffer in their succeeding tiles. After two cycles, partial-sums \ding{183}, \ding{186}, and \ding{189} are generated in the second tiles in their groups, which are then added with the stored partial-sums \ding{182}, \ding{185}, and \ding{188}, respectively. The computing process is similar to the dataflow without weight duplication except that $K$ group-sums are computed in $K$ columns. 

\textbf{The group-sum dataflow}. A group-sum dataflow with weight duplication aims to reduce the data moving distance by adding up group-sums locally. Blue and pink arrows indicate the group-sum dataflow in \autoref{fig:conv_dataflow_dup} (a). We take the convolution results $\mathbf{Q}$ and $\mathbf{T}$ for instances. Based on the input dataflow and partial-sum dataflow, group-sums $\mathbf{Q_1}$ and $\mathbf{Q_2}$ are computed in the first two columns of a block. Thus $\mathbf{Q_1}$ is transmitted from the first column to the adjacent column when it reaches the sixth tile. $\mathbf{Q_2}$ is stored in the Rofm buffer waiting for $\mathbf{Q_1}$. $\mathbf{Q_2}$ will be popped out by instruction when Rofm receives $\mathbf{Q_1}$. They will be added up before transmitting along the second column. The pink arrows represent another situation that the second group-sum is not waiting in Rofm. Based on the input dataflow, $\mathbf{T_1}$ and $\mathbf{T_2}$ are computed in the same column but in two turns of inputs. When $\mathbf{T_1}$ reaches the sixth tile, $\mathbf{T_2}$ is not ready yet. Therefore, $\mathbf{T_1}$ will be stored in the Rofm buffer to wait for the fifth row of data to flow into the tiles. Once $\mathbf{T_2}$ is computed in the sixth tile, $\mathbf{T_1}$ and $\mathbf{T_2}$ are added up to get $\mathbf{T_1} + \mathbf{T_2}$. The group-sum dataflow greatly reduces the data moving distance and frequency.

\subsection{Pooling Dataflow}
\begin{figure}[ht]
    \centering  
    \includegraphics[width=0.37\textwidth]{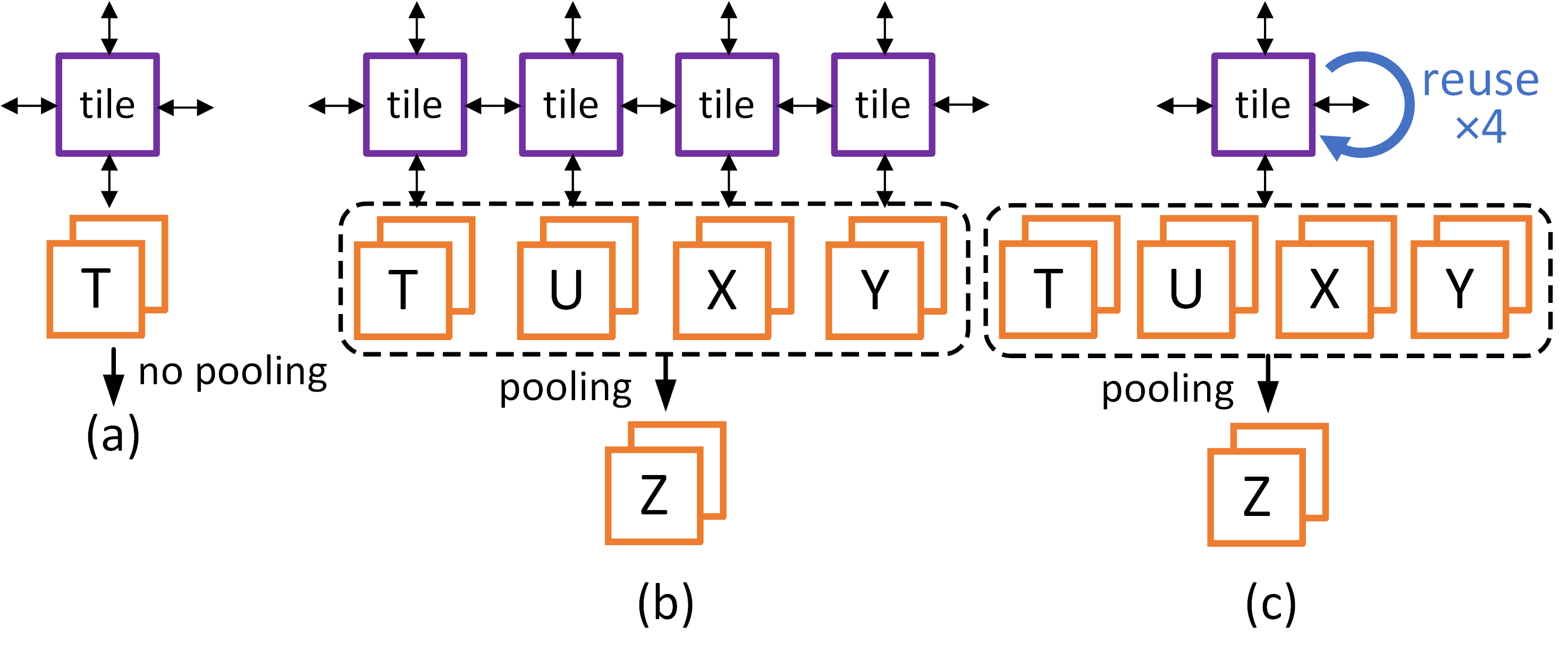}
    \caption{Output in the last tile: weight duplication or block reuse scheme is used to deal with pooling layer.}  
    \label{fig:pooling}
\end{figure}

The computation of CONV and FC layers is processed within Domino blocks, while the computation of the pooling layer is performed during data transmission between blocks. If a pooling layer follows a CONV layer, with pooling filter size $K_p = 2$ and pooling stride $S_p = 2$, every four activation results produce a pooling result. Weight duplication situation is shown in \autoref{fig:pooling} (b), in every cycle a block produces four activation results $\mathbf{T}$ to $\mathbf{Y}$. When transmitting across tiles, the data are compared, and the pooling result $\mathbf{Z}$ is computed. \autoref{fig:pooling} (c) shows the block reuse case that activation results are computed and stored in the last tile. A comparison is taken when the next activation result is computed. The Rofm outputs a pooling result $\mathbf{Z}$ once the comparison of the pooling filter is completed. In this case, the computation frequency before pooling layers is $4\times$ higher than the succeeding blocks.

\section{Instructions and Workflow}
\label{instruction}
Domino is a highly distributed and decentralized architecture and adopts self-controlled instructions, which are stored in each Rofm. The reason is that Domino tries to reduce the bandwidth demand for transmitting data or instructions through whole NoC, and avoid the delay and skew of long-distance signals; meanwhile, localized instructions retain flexibility and compatibility for different DNNs to be processed.

\subsection{Instruction Set}
A set of instructions is designed for computing and transmitting data automatically and flexibly when dealing with different DNNs. In Domino, Rofm is responsible for sending packets and accumulating partial-sums and group-sums correctly. Because our packets transmitted through NoC only contain payloads of DNN without any other auxiliary information (i.e., a message header, a tail, etc.), there must be an intuitive way to control the behavior of each Rofm on the chip. Consequently, we define an instruction set for Domino.

The instruction consists of several fields, each of which represents a set of control words for relative ports or buffers, as well as the designed actions, as shown in \autoref{tab:instruction}. The instruction length is 16 bits, including four distinct fields: an Rx field, a Function field, a Tx field, and an opcode field. The Rx field takes responsibility for receiving packets from different ports. The Function field contains sum and buffer control in C-type, activation function, pooling, and FC layer control in M-type. The Tx field governs the transmission of one Rofm to four ports. There are two types of opcodes bearing two usages. C-type (stands for convolution type) denotes the instruction is served for controlling the process during convolution computation, while the M-type is for miscellaneous operations other than convolution, such as activating and pooling.

\begin{table}[h]
 
\begin{center}
\begin{tabular}{p{0in}p{0.05in}p{0.05in}p{0.05in}p{0.05in}p{0.05in}p{0.05in}p{0.05in}p{0.05in}l}
 
\multicolumn{1}{l}{\instbit{15}} &
\multicolumn{1}{r}{\instbit{11}} &

\multicolumn{1}{l}{\instbit{10}} &
\multicolumn{1}{l}{\instbit{7}} &

\multicolumn{1}{l}{\instbit{6}} &
\multicolumn{1}{l}{\instbit{5}} &

\multicolumn{1}{l}{\instbit{4}} &
\multicolumn{1}{l}{\instbit{1}} &
\multicolumn{1}{r}{\instbit{0}} &  \\
\cline{1-9}

\multicolumn{2}{|c|}{Rx Ctrl.} &
\multicolumn{2}{c|}{Sum} &
\multicolumn{2}{c|}{Buffer} &
\multicolumn{2}{c|}{Tx Ctrl.} &
\multicolumn{1}{c|}{Opc.} &  C-type \\
\cline{1-9}

\multicolumn{2}{|c|}{Rx Ctrl.} &
\multicolumn{4}{c|}{Func} &
\multicolumn{2}{c|}{Tx Ctrl.} &
\multicolumn{1}{c|}{Opc.} &  M-type \\
\cline{1-9}

\end{tabular}
\end{center}
 
\caption{The instruction format for Domino. There are two basic types of instructions generated by the compiler to handle dataflow correctly.}
\label{tab:instruction}
\end{table}

\subsection{Schedule Table}
In Domino, instructions in a Rofm should fit both intra- and inter- block dataflows to support the ``computing-on-the-move'' procedure. The compiler generates the instruction table and configuration information for each tile based on the initial input data and the DNN structure. The highly distributed and local-controlled structure avoids the extra instruction transmission and external control signals when processing DNNs. 

After cycle-accurate analyses and mathematical derivation, instructions reveal an attribute of periodicity. During the convolution computation, C-type instructions are fetched from the schedule table and executed periodically. The period $p$ ($p=2(P+W)$) is related to IFM. Within a period, actions are fixed to the given IFM and DNN configurations. Furthermore, every port's behavior exhibits a period of $p$ with a different beginning time. The control words for each port are stored in Rofm, and they are generated based on the assumption that the convolution stride is one. When the convolution stride is not one, the compiler will shield certain bit in control words to ``skip'' some actions in the corresponding cycles to meet the required dataflow controlling. When a Rofm is mapped and configured to process the last row of a layer in a CNN, it will generate activation and pooling instructions. Its period is related to pooling stride, $p=2S_p$.

Once instructions for each Rofm are received and stored, the Rofm is configured to prepare for computation. For a given Rofm, when a clock cycle begins, a counter provides an index to Rofm to fetch corresponding control words periodically. When the execution of an instruction ends, the state of the current Rofm will be updated. 

\subsection{Workflow}
\begin{figure*}[htbp]
\centering
    \includegraphics[width=1\textwidth]{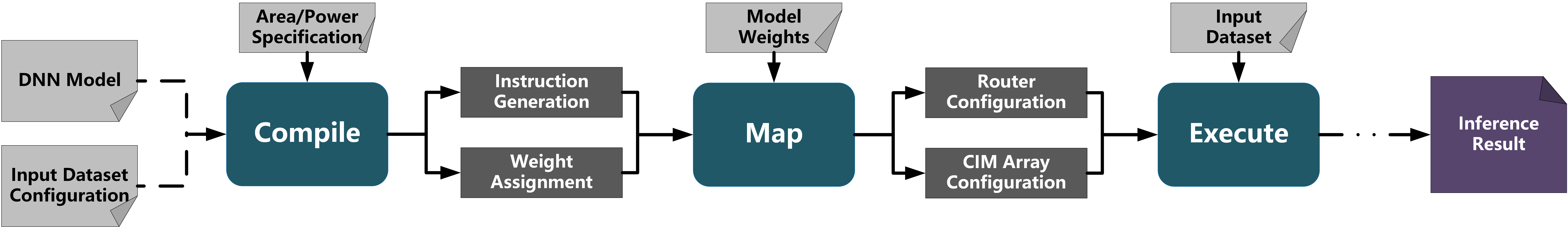}
    \caption{The workflow of Domino. Domino takes network model, dataset information, and area or power specification for compilation. The generated schedule tables and weights are mapped into routers and CIM arrays respectively. After initialization, Domino begins to execute instructions and infer.}  
    \label{fig:workflow}
\end{figure*}
After elaborated introduction and illustration of Domino technical details, the overall workflow and processing chain are naturally constructed and revealed. \autoref{fig:workflow} shows how Domino works in a sequence to process DNN. These building blocks cover both ideas and implementations mentioned above and integrate them into a whole architecture.

\section{Evaluation}
\label{evaluation}
This section evaluates Domino's characterization and performances in detail, including energy efficiency and throughput. Meanwhile, we compare Domino against other state-of-the-art CIM-based architectures on several prevailing types of CNNs to show the proposed architecture and dataflow advantages.

\subsection{Methodology}
First, we specify the configurations of our Domino under test; second, we describe the experiment setup and benchmarks; then we determine the normalization methods.
 
\begin{scriptsize}
    \begin{table}[ht]
    \footnotesize
    \begin{center}
    \begin{tabular}{|c|c|c|c|}
    \hline
        \textbf{Component} & \textbf{Descript.} & \textbf{Energy/Compo.} & \textbf{Area ($\upmu$m$^{2}$) }\\
        \hline
        ADC & 8b$\times$256 & 1.76 pJ & 351.5 \\ 
        \hline
        Integrator & 8b$\times$256 & 2.13 pJ  & 568.3  \\ 
        \hline
        Crossbar cell & 8b & 32.9 fJ  & 1.62  \\ 
 
        \hline
        \textbf{PE total}& \makecell[c]{$N_c$$ \times $$N_m$} & 48.1 fJ/MAC & 341632 \\ 
        \hline
        \hline
        Buffer & 256B$\times$1  & 281.3 pJ& 826.5 \\
        \hline

        Ctrl. circ. & 1 & 4.1 pJ & 1400.6 \\       \hline
        \textbf{Rifm total} & \multicolumn{2}{c|}{-}  & 2227.1  \\
        \hline        
        \hline

        Adder & 8b$\times$8$\times$2 & 0.03 pJ/8b & 0.07 \\
        \hline
        Pooling & 8b$\times$8 & 7.6 fJ/8b & 34.06 \\
        \hline
        Activation & 8b$\times$8 & 0.9 fJ/8b & 7.07 \\
        \hline
        Data buffer & 16KiB  &  281.3 pJ & 52896 \\
        \hline
        Sched. Table & 16b$\times$128  &  2.2 pJ/16b & 826.5 \\
        \hline
        Input buffer & 64b$\times$2  & 17.6 pJ& 51.7  \\
        \hline
        Output buffer & 64b$\times$2  & 17.6 pJ& 51.7  \\
        \hline
        Ctrl. circ.  & 1 &28.5 pJ & 2451.2 \\
        \hline 
        \textbf{Rofm total} &\multicolumn{2}{c|}{-} & 53867.1 \\
        \hline
        \hline
        \textbf{Tile total}  & \multicolumn{2}{c|}{-}  & 0.398 mm$^{2}$\\
        \hline
        \end{tabular}
    \caption{The configuration information summary for Domino under evaluation.}
    \label{tab:ciim_config}
    \end{center}
    \end{table}
\end{scriptsize}

\subsubsection{Configuration}
The configuration of Domino and its tile is displayed in \autoref{tab:ciim_config}. Domino adopts silicon-proven ADC in \cite{ADC}, silicon-proven SRAM array in \cite{sram}, and the transmission parameters in \cite{Noxim1}. The rest of components are simulated by spice or synthesized by Synopsys DC with a 45 nm CMOS Process Design Kit (PDK). Domino runs at a step frequency of 10 MHz for instruction updating. The data transmission frequency is 640 MHz, and the data width is 64 bits. The supply voltage is 1 V. ReRAM's area information is normalized from the silicon-proven result in \cite{ISSCC_2020_Liu}. In the initialization stage, each ReRAM cell in a PE is programmed with a 1-bit weight value. Eight ReRAM cells make up an 8-bit weight. At the same time, each Rifm and Rofm is configured, and the schedule tables pre-load instructions. Once the configuration of Domino is finished, the instructions are ready for execution, and Domino will commence processing DNNs when the clock cycle starts.

\subsubsection{Experiment Setup}
We select several types of CIM accelerator architectures as the reference, including three silicon-proven architectures and another four simulated architectures. The performance data and the configuration parameters of the baselines are taken from their papers. The following three experiments are conducted: (1) running different architectures with the benchmark DNNs to compare the power efficiency; (2) evaluating the throughput of Domino under different DNNs and input; and (3) estimating PE utilization with various DNN models and PE size. 

\subsubsection{Benchmarks}
We evaluate Domino using representative DNN models like VGG-11\cite{ISSCC_2021_Jia}, VGG-16, VGG-19\cite{vggnet}, ResNet-18, and ResNet-50\cite{resnet} as the model benchmarks. The datasets CIFAR-10\cite{CIFAR-10} and ImageNet\cite{imagenet} are chosen to be dataset benchmarks for evaluation. Then we run benchmarks consistent to each comparison object respectively.

\subsubsection{Normalization}
A variety of CIM models lead to discrepancies of many attributes that need normalization and scaling. To make a fair comparison, we normalize technology nodes and supply voltage of digital circuits in each design to 45 nm and 1 V according to the equations given in \cite{node-scaling}. The activations' and weights' precision is scaled to 8-bit as well. The supply voltage of analog circuits is also normalized to 1 V.

\subsection{Performance Results}

\begin{table*}[htb]
    \begin{center}
    \begin{threeparttable}
    \scriptsize
    \begin{tabular}{|c|c|c|c|c|c|c|c|c|c|c|c|c|}
    \hline
        Dataset & \multicolumn{4}{c|}{CIFAR-10}  & \multicolumn{8}{c|}{ImageNet} \\
        \hline
        Model  & \multicolumn{2}{c|}{VGG-11} & \multicolumn{2}{c|}{ResNet-18} & \multicolumn{3}{c|}{VGG-16} & \multicolumn{3}{c|}{VGG-19} & \multicolumn{2}{c|}{ResNet-50} \\
        \hline
        Architecture & \cite{ISSCC_2021_Jia} & Ours & \cite{ISSCC_2020_Yue} & Ours & \cite{ISSCC_2021_Yoon}\textcolor{gray}{\tnote{*1}}  & \cite{MAERI} & Ours & \cite{AtomLayer} & \cite{CASCADE} & Ours & \cite{timely} & Ours\\
        \hline
        \makecell[c]{CIM type} 
        & SRAM  & ReRAM  & SRAM    & ReRAM & ReRAM  &  n.a. & ReRAM & ReRAM  & ReRAM   & ReRAM & ReRAM & ReRAM \\
        \hline
        \makecell[c]{Tech. (mm)} 
        & 16    & 45     & 65      & 45    & 40     &  28   & 45    & 32     & 65      & 45    & 65    & 40 \\
        \hline
        \makecell[c]{VDD (V)} 
        & 0.8   & 1      & 1       & 1     & 0.9    & n.a.  & 1     & 1      & 1       & 1     & 1.2   & 1 \\
        \hline
        \makecell[c]{Frequency (MHz)} 
        & 200   & 10     & 100     & 10    & 100    & 200   & 10    & 1200   & 1200    & 10    & 40    & 10\\
        \hline
        \makecell[c]{Act. \& W. precision} & 4 & 8 & 4 & 8  & 8  & 16 & 8 & 16 &16  & 8 & 8 & 8  \\
        \hline
        \makecell[c]{\# of CIM array} 
        & 16    & 900    & 4      & 900   & 1      & n.a.   & 2500   & 2560   & 6400   & 2500  & 20352  & 900  \\
        \hline
        \makecell[c]{CIM array size (kb)} 
        & 288   & 512    & 16     & 512   & 64     & n.a.   & 512     & 256    & 4     & 512     & 256   & 512  \\
        \hline
        \makecell[c]{Area (mm$^2$)} 
        & 25    & 358.2  & 9      & 358.2 & 9      & 6      & 995     & 5.32   & 0.99  & 995     &  91   & 358.2   \\
        \hline
        \makecell[c]{Tapeout/Simulation} 
        & T     & S      & T      & S     & T      &  S     & S       & S      & S       & S     & S     & S  \\
        \hline
        \makecell[c]{Exec. time (us)} 
        & 128   & 129.5  & 1890   & 203.5 & 0.67s  & n.a.   & 3471    & 6920   & n.a.    & 3557  &  n.a.  & 2397  \\
        \hline
        \makecell[c]{CIM energy (uJ)} 
        & 11.47 & 36.74  & 36.11  & 26.44  & 3700  & 0      & 744.1   & 25000  & 13000   & 944.3  & 130.2  & 168.3 \\
        \hline
        \makecell[c]{On-chip data\\ moving energy (uJ)} 
        & 0.16  &  2.63  & n.a.   &  3.89  & n.a.  & 36741  & 46.39   & n.a.   & n.a.    & 52.81   & 0.64  & 16.97  \\
        \hline
        \makecell[c]{On-chip memory\\ energy (uJ)} 
        & 5.11  &  25.41 & n.a.   & 24.21  & n.a.  & 27556  & 446.4   & 2310   & 2180    & 508.1   & 71.22    & 115.41 \\
        \hline
        \makecell[c]{Other comp. (uJ)\textcolor{gray}{\tnote{*2}}} 
        & 2.67  & 0.48   &   n.a. & 0.46   &  n.a. & 50519  & 8.41    &   n.a. & 1040    & 9.59    & 44.96    & 1.68 \\
        \hline
        \makecell[c]{Off-chip data\\ accessing energy (uJ)} 
        & n.a.  &  0     & 123.89 &   0    & 3690  & n.a.   & 0       & 5650   & 4000    & 0      & 85.98     & 0 \\
        \hline
        \makecell[c]{Power (W)} 
        & 0.17 & 40.78  & 0.005   & 34.38  & 0.011 & 0.54   & 15.89   & 4.8    & 0.003   & 19.33   & 166      & 30.84  \\
        \hline
        \makecell[c]{CE (TOPS/W)} 
        & 71.39 & 23.41  & 6.91  & 19.99   & 4.15   & 0.27   & 24.84   & 0.68   & 3.55   & 25.92  & 21     & 23.14 \\
        \hline
        \makecell[c]{Normalized\textcolor{gray}{\tnote{*3}}\\ CE (TOPS/W)} 
        & 9.53  & 23.41  & 2.82  & 19.99   & 9.24   & 0.36   & 24.84   & 2.73   & 12.98  & 25.92  & 22.46  & 23.14 \\
        \hline
        \makecell[c]{Throughput (TOPS)} 
        & 12.14 & 954.66 & 0.036 & 687.26  & 0.046  & 0.15   & 394.7   & 3.28   & 0.1    & 501    & 3488   & 713.6 \\
        \hline
        \makecell[c]{Throughput\\(TOPS/mm$^2$)} 
        & 0.49  & 2.67   & 0.004 & 1.92    & 0.005  & 0.025  & 0.4     & 0.62   & 0.10   & 0.5    & 38.33  & 1.99 \\
        \hline
        \makecell[c]{inferences/s}
        & 7815  & 6.25E5 &  n.a. & 6.25E5  & n.a.   & n.a.   & 1.28E4  &   n.a. &   n.a.  & 1.28E4 & n.a.  & 1.02E5  \\
        \hline
        \makecell[c]{Accuracy(\%)} 
        & 91.51 & 89.85 & 91.15 &  91.57  & 46     &  n.a.  &   70.71  & n.a.   &  n.a.  & 72.38   & 76    & 74.89 \\
        \hline
    \end{tabular}
    \begin{tablenotes}
        \item \textcolor{gray}{*1} \textcolor{gray}{Adapted and normalized from average statistics.} \textcolor{gray}{*2} \textcolor{gray}{Including CNN related computation like partial sum accumulation, activation functions, pooling operations, etc.} \textcolor{gray}{*3} \textcolor{gray}{Voltage normalized to 1 V, precision normalized to 8-bit and digital circuit normalized to 45 nm referring to \cite{node-scaling}.}
    \end{tablenotes}
    \caption{Domino's evaluation results and comparison under different DNN models.}
    \label{tab:domino_performance}
    \end{threeparttable}
    \end{center}
\end{table*}

We evaluate Domino's Computational Efficiency (CE), power dissipation, energy consumption, throughput, and execution time for each experiment environment in benchmarks, and compare them with other architectures. 

\subsubsection{Overall Performance}
We evaluate Domino on various DNN models to prove that its performance can be generalized to different model patterns. \autoref{tab:domino_performance} shows Domino's system execution time, energy consumption, and computational efficiency when it runs benchmarks with specified configurations. The time and energy spent on initialization and compilation are not considered. We break down energy into five parts: CIM, on-chip data moving, on-chip data memory, other computation, and off-chip data accessing. The first one is the energy consumed in PEs related to MAC operations. The latter four are the energy of peripheral circuits. Domino achieves a peak CE of 25.92 TOPS/W when running VGG-19 with ImageNet, and a valley CE of 19.99 TOPS/W running ResNet-18 with CIFAR-10. It denotes that the larger the DNN model size, the better CE it achieves. The peripheral energy in VGG-19 only occupies one-third of the total energy. The throughput of Domino varies with input data size and DNN model. VGG and ResNet models with a small dataset could achieve 6.25E5 inferences/s high throughput. VGG-19 with ImageNet dataset still has a throughput of 1.276E4 inferences/s. The high throughput and low peripheral energy of Domino are benefited from the pipelined ``computing-on-the-move'' dataflow that fully utilizes the data locality.

\subsubsection{Computational Efficiency}

The comparison between Domino and other state-of-the-art CIM architectures and DNN accelerator is summarized in \autoref{tab:domino_performance}. The data in the table are either taken or calculated based on the results provided in their papers. The silicon-proven designs usually have low throughput and power consumption due to the small chip size. The typical bit width is only 4-bit because analog computing is difficult to achieve a high signal-to-noise ratio. It can be seen between the CIM computing efficiency and the system computing efficiency that energy consumed by the peripheral circuits for data moving and memory accessing may still dominate the total power consumption in most of the design. Benefited from a 16 nm technology node and tailored DNN model, Jia \cite{ISSCC_2021_Jia} could achieve a smaller gap between the CIM computing efficiency (121 TOPS/W) and the system computing efficiency (71.39 TOPS/W) at 4-bit configuration.

\begin{figure}[ht]
    \includegraphics[width=0.48\textwidth]{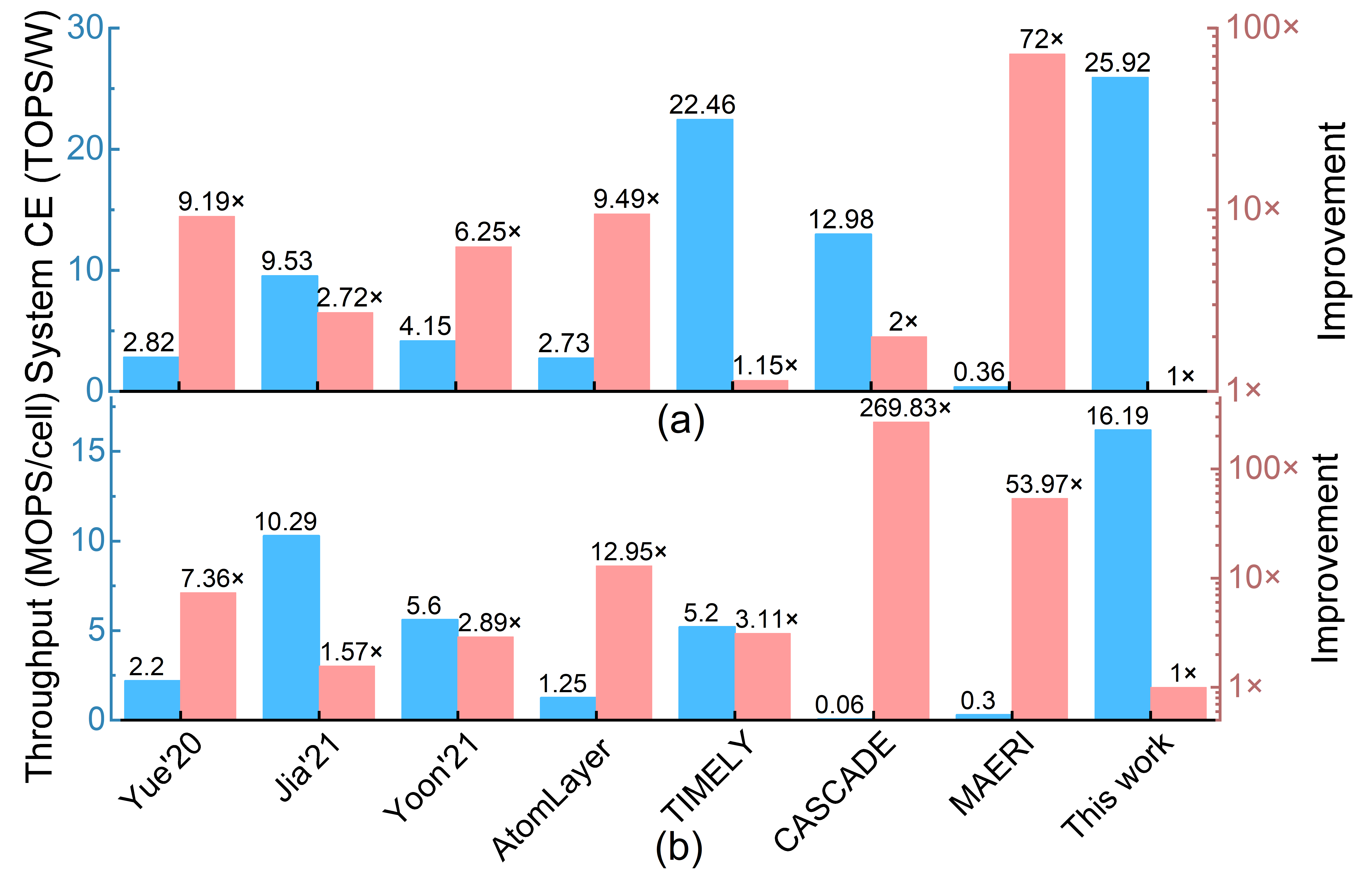}
    \caption{Performance comparisons with other architectures (a) System computational efficiency after normalization; (b) Throughput normalized to each 8-bit crossbar array cell.}  
    \label{fig:performance}
\end{figure}

To make a fair comparison among different designs, all circuits are normalized to 1 V, 8-bit and digital components are further normalized to 45 nm. From \autoref{fig:performance} we can see that Domino achieves the highest system CE (25.92 TOPS/W), which is 1.15-9.49$\times$ higher than the CIM counterparts. Moreover, Domino has a lower percentage of memory accessing, data transmitting energy (in \autoref{tab:domino_performance}). The energy breakdown shows that Domino effectively reduces peripheral energy consumption, which is a vital factor in the overall system performance. The data locality and ``computing-on-the-move'' dataflow are very efficient in reducing the overall energy consumption for DNN inference.

As for DNN accelerators, Domino overwhelms MAERI\cite{MAERI} on CE, which is the representative of conventional digital architecture. The superiority comes not only from the use of CIM but also from ``computing-on-the-move'' dataflow.

\subsubsection{Throughput}
Domino has a distinguished advantage over other architectures in terms of throughput. The existing architectures need to store feature maps and partial-sums to external memory to support the complex dataflow. Furthermore, some designs also have to access external memory to update weights for different layers. In Domino, the weights of the DNN are all stored in crossbar arrays in the initial configuration. Layer synchronization with weight duplication and block reuse scheme is further proposed to maximize the parallel computing speed of inference.

The chip size and the number of memory cells greatly affect the overall throughput of a system, and different memory type has different memory size (i.e., ReRAM and SRAM). As shown in \autoref{tab:domino_performance}, the area normalized throughput of \cite{ISSCC_2020_Yue} is 0.004 TOPS/mm$^2$ due to the small array size and 65 nm technology node. With a larger crossbar array size and 16 nm technology node, \cite{ISSCC_2021_Jia} achieves 0.486 TOPS/mm$^2$. Our proposed scheme can achieve up to 2.67 TOPS/mm$^2$ throughput, which is better than the state-of-the-art schemes except \cite{timely}. The high density of \cite{timely} is due to the transistor-less crossbar array estimated on 20nm technology node, which often suffers from the write sneak path issue and is very difficult to program the ReRAM cells to the desired state. Therefore, we normalize the throughput to an 8-bit cell, which reflects both throughput and PE utilization. As shown in \autoref{fig:performance} (b), our design achieves $16.19$ MOPS/8-b-cell throughput, which is 3.10$\times$ higher than TIMELY and 270$\times$ higher than CASCADE. Although \cite{ISSCC_2021_Yue} supports zero-skipping and implements a ping-pong CIM for computing while refreshing, and \cite{ISSCC_2021_Jia} provides various mapping strategies to improve the throughput, Domino still outperforms them by 7.36$\times$ and 1.57$\times$, respectively. The high throughput of Domino comes from the synchronization, the weight pre-loading, the data locality, and the pipelined dataflow to reduce the computing latency while maximizing the parallelism.

\subsubsection{Utilization Rates}
\begin{figure}[htbp]
    \centering
    \includegraphics[width=0.45\textwidth]{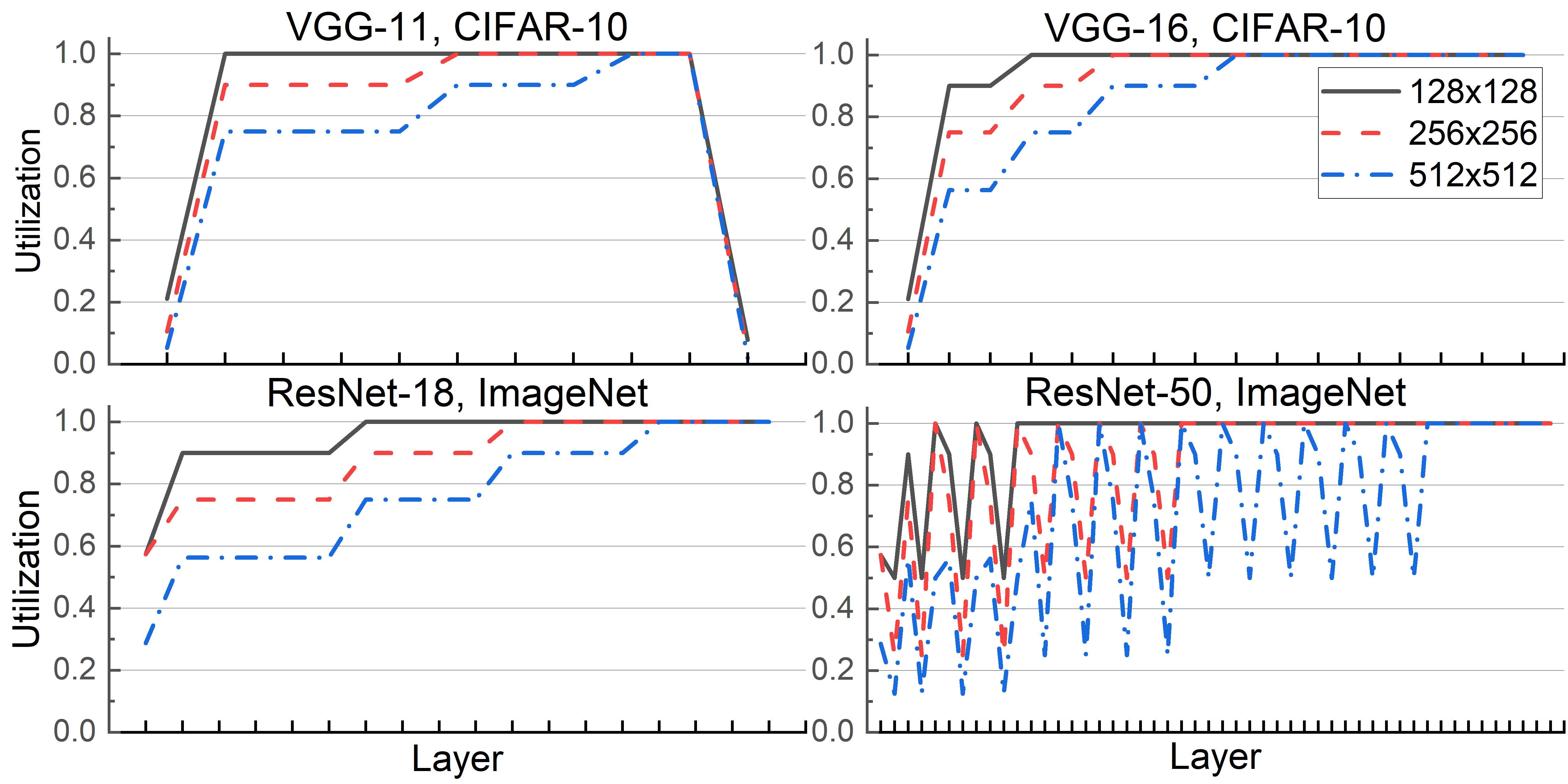}
    \caption{Utilization rates over all layers in four DNN models considering three PE crossbar configurations.}  
    \label{fig:Utilization}
\end{figure}

The size of the crossbar array greatly affects the CIM power efficiency and cell utilization in PEs. \autoref{fig:Utilization} depicts the crossbar array utilization over all layers in four neural network models: VGG-11, VGG-16, ResNet-18, and ResNet-50 with three crossbar array configurations ($128\times 128$, $256\times 256$, and $512\times 512$). Though the mapping strategies in Domino have improved cell utilization in PEs, they still have low utilization in the first few layers because the input and output channels are much less than the side length of the crossbar array. The average utilization of four models is 98\%, 96\%, 93\%, and 92\% when using a 128$\times$128 crossbar array. The utilization is reduced to 90\%, 89\%, 86\% and 79\% using a 256$\times$256 crossbar array, and 75\%, 76\%, 67\% and 54\% with a 512$\times$512 crossbar array. Lower utilization in ResNet comes from its architecture that layers with small channels are prevalent. Though a smaller crossbar array has higher utilization, it sacrifices the CIM computing efficiency, which is 31.4 TOPS/W, 41.58 TOPS/W, and 49.38 TOPS/W at these three configurations. Therefore, it is a balance between utilization and computational efficiency.

\section{Related Work}
\label{related}
Various CIM architectures have been proposed to adopt different designs, computing strategies, and architectures. SRAM- or DARAM-based CIM chips usually need to update weights by accessing off-chip memory \cite{ISSCC_2021_Jia,ISSCC_2021_Chen,ISSCC_2021_Yue,ISSCC_2021_Guo}, resulting in high energy consumption and latency. \cite{ISSCC_2021_Jia} utilizes 1152 input rows to fit a 3$\times$3 convolution filter, which will face low cell utilization in other DNN configurations. ISAAC\cite{ISAAC} is a ReRAM-based accelerator that uses ReRAM crossbars to store weights and eDRAM to store feature maps, which are organized by routers and H-trees. Pipelayer\cite{PipeLayer} stores both filters and feature maps in ReRAM crossbars and computes them in the pipelined dataflow. The interface between analog domain and digital domain also contributes to high power consumption. \cite{zhang2020regulator} proposes a binary input pattern for CIM, instead of a high power consumption DAC, to reduce the interface energy. In this way, both computing efficiency and reliability are improved. CASCADE\cite{CASCADE} applies the significance of the bit lines in the analog domain to minimize A/D conversions, thus enhancing power efficiency. TIMELY\cite{timely} also focuses on analog current adder to increase the CIM block size, TDC and DTC to reduce the power consumption, and input data locality to reduce data movement. However, the uncertainty caused by the clock jitter and the time variation caused by the complex RRAM crossbar network will greatly reduce the ENOB of the converters. Moreover, the large sub-chip and weight duplication inside the crossbar array also significantly reduce the utilization of the CIM arrays. Till now, most of the designs don't provide on-chip control circuits to maintain the complicated dataflow. Therefore, the feature maps or partial-sums have to be stored on the external memory and controlled by additional processors to transform the feature maps' flow and synchronize the computing latency of different data paths. Accessing feature maps and partial-sums have been the leading energy consumption in most designs. Reducing the data movement energy has been one of the most important research topics to design CIM processors.

\section{Conclusion}
\label{conclusion}
This paper has presented a tailored Network-on-Chip architecture called Domino with highly localized inter- and intra-memory computing for DNNs. The key contribution and innovation can be concluded as follows: Domino (1) changes conventional NoC tile structure by using two dual routers for different usages aimed at DNN processing, and enabling architecture to substitute PE for different types of CIM arrays, (2) constructs highly distributed and self-controlled processor architecture, proposes efficient matching dataflow to perform ``computing-on-the-move'', and (3) defines a set of instructions for routers, which is executed periodically to process DNNs. Benefiting from such design, Domino has unique features with respect to (1) 
eliminating data access to memory during one single inference, (2) minimizing data movement on-chip, (3) achieving high computational efficiency and throughput. Compared with the competitive architectures, Domino has achieved 1.15-to-9.49$\times$ power efficiency improvement and improved the throughput by 1.57-to-12.96$\times$ over several current advanced architectures \cite{TCAS-II}\cite{ISSCC_2020_Zhang}\cite{ISSCC_2020_Liu}.

\bibliographystyle{IEEEtranS}
 
\bibliography{refs}

\end{document}